\documentclass[11pt]{article}
\usepackage{amsmath,amssymb,color,epsfig,cite}
\usepackage{graphicx}
\usepackage{subfigure}
\usepackage{setspace}
\usepackage{amsfonts}
\newcommand{\hoch}[1]{$\, ^{#1}$}

\makeatletter
\@addtoreset{equation}{section}
\makeatother

\newcommand{\be}{\begin{equation}}
\newcommand{\ee}{\end{equation}}
\newcommand{\bea}{\setlength\arraycolsep{2pt} \begin{eqnarray}}
\newcommand{\eea}{\end{eqnarray}}
\newcommand{\nn}{\nonumber}

\def\ft#1#2{{\textstyle{\frac{\scriptstyle #1}{\scriptstyle #2} } }}
\def\fft#1#2{{\frac{#1}{#2}}}

\def\0{{\sst{(0)}}}
\def\1{{\sst{(1)}}}
\def\2{{\sst{(2)}}}
\def\3{{\sst{(3)}}}
\def\4{{\sst{(4)}}}
\def\5{{\sst{(5)}}}
\def\6{{\sst{(6)}}}
\def\7{{\sst{(7)}}}
\def\8{{\sst{(8)}}}
\def\9{{\sst{(9)}}}

\def\sst#1{{\scriptscriptstyle #1}}

\begin{document}

%\begin{flushright}
%\hfill{MI-TH-1762}

%\end{flushright}

\begin{center}
{\large {\bf Holographic Studies of The Generic Massless Cubic Gravities
}}

\vspace{10pt}
Yue-Zhou Li\hoch{1\dag}

\vspace{15pt}

{\it \hoch{1}Center for Joint Quantum Studies and Department of Physics,\\
School of Science, Tianjin University, Tianjin 300350, China}

\vspace{30pt}

\underline{ABSTRACT}
\end{center}
We consider the generic massless cubic gravities coupled to a negative bare cosmological constant mainly in $D=5$ and $D=4$ dimensions, which are Einstein gravity extended with cubic curvature invariants where the linearized excited spectrum around the AdS background contains no massive modes. The generic massless cubic gravities are more general than Myers quasi-topological gravity in $D=5$ and Einsteinian cubic gravity in $D=4$. It turns out that the massless cubic gravities admit the black holes at least in a perturbative sense with the coupling constants of the cubic terms becoming infinitesimal. The first order approximate black hole solutions with arbitrary boundary topology $k$ are presented, and in addition, the second order approximate planar black holes are exhibited as well. We then establish the holographic dictionary for such theories by presenting $a$-charge, $C_T$-charge and energy flux parameters $t_2$ and $t_4$. By perturbatively discussing the holographic R\'enyi entropy, we find $a$, $C_T$ and $t_4$ can somehow determine the R\'enyi entropy with the limit $q\rightarrow 1$, $q\rightarrow 0$ and $q\rightarrow \infty$ up to the first order, where $q$ is the order of the R\'enyi entropy. For holographic hydrodynamics, we discuss the shear-viscosity-entropy-ratio and find that the patterns deviating from the KSS bound $1/(4\pi)$ can somehow be controlled by $((c-a)/c,t_4)$ up to the first order in $D=5$, and $((\mathcal{C}_T-\tilde{a})/\mathcal{C}_T,t_4)$ up to the second order in $D=4$, where $\mathcal{C}_T$ and $\tilde{a}$ differ from $C_T$-charge and $a$-charge by inessential overall constants.

\vfill {\footnotesize \hoch{\dag}liyuezhou@tju.edu.cn}

\pagebreak

\tableofcontents
\addtocontents{toc}{\protect\setcounter{tocdepth}{2}}

%%%%%%%%%%%%%%%%%%%%%%%%%%%%%%%%%%%%%%%%

\newpage
%%%%%%%%%%%%%%%%%%%%%%%%%%%%%%%%%%%%%%%%
\section{Introduction}
\label{sec:intro}
Einstein gravity extended with higher-order curvature invariant terms has acquired considerable attentions, especially in the context of AdS/CFT \cite{Maldacena:1997re,Gubser:1998bc,Witten:1998qj}. Coupled with a bare negative cosmological constants, anti-de Sitter (AdS) vacua can automatically arise with an effective AdS radius in higher-order gravity theories, suggesting higher-order gravities can serve as holographic models to investigate a variety of properties for some dual conformal field theory (CFT)\footnote{In this paper, the bulk dimension is written as $D$, the dimension of the boundary CFT is written as $d=D-1$.}. However, in general, such AdS vacua are unstable and their perturbative excitations would contain the extra ghosty massive spin-$2$ mode and massive scalar mode (see, e.g. \cite{Tekin:2016vli,Bueno:2016xff,Bueno:2016ypa} for more exhaustive and comprehensive discussions). Removing the ghost mode is compulsive, otherwise the dual CFT would not be unitary. On the other hand, the decoupling of the massive scalar mode is the necessary condition for holographic $a$-theorem \cite{Li:2017txk}.\footnote{However, decoupling of the massive scalar mode is not the sufficient condition for $a$-theorem. $a$-theorem itself requires more constraints, see e.g. \cite{Li:2017txk}.} With both the massive spin-$2$ mode and massive scalar mode being decoupled, the linearized spectrum around the AdS vacua contains only the graviton modes and the resulting theory is referred as the massless gravity \cite{Li:2018drw}. The possibly simplest examples of massless gravity are the Gauss-Bonnet combination and more generic Lovelock gravities \cite{ll}. Essentially, massless gravities are likely to have well-defined CFT dual, and hence it is of great importance and interests to study the effect of the coupling constants involved in higher-order terms of massless gravities on various CFT properties.

The first step to understand the holographic aspects of a gravity theory is to establish the holographic dictionary in which, by applying the holographic renormalization scheme  \cite{deBoer:2000cz,Bianchi:2001kw,Skenderis:2002wp}, the conformal anomaly \cite{Duff:1977ay,Duff:1993wm}, two-point function and three-point function of energy-momentum tensor \cite{Osborn:1993cr,Erdmenger:1996yc,Coriano:2012wp} shall be revealed holographically, e.g. \cite{Henningson:1998gx,Henningson:1998ey,Imbimbo:1999bj,Nojiri:1999mh,Liu:1998bu,Johansson:2012fs,Ghodsi:2014hua,Li:2018drw,Arutyunov:1999nw}. The conformal anomaly, two-point function and three-point function respectively have uniquely determined structures that are shaped by conformal invariance, and the properties of a CFT are attached to the parameters in them, i.e. $a$-charge that measures the massless degree of freedom and indicates the property of RG flow \cite{zamo,Cardy:1988cwa,Komargodski:2011vj}, $C_T$ coefficient (or equivalently, $c$-charge in $d=4$) that can determine the two-point function and $\mathcal{A}$, $\mathcal{B}$, $\mathcal{C}$ that can determine the three-point function \cite{Osborn:1993cr,Erdmenger:1996yc,Coriano:2012wp}. According to the conformal collider thought experiment proposed in \cite{Hofman:2008ar}, three-point function can be equivalently described by the one-point function of the energy flux excited by some local excitations, and the parameters $t_2$ and $t_4$ appearing there play the similar role as $\mathcal{A}$, $\mathcal{B}$ and $\mathcal{C}$.

The simplest example for holographic dictionary is the case of quadratic order where Gauss-Bonnet gravity is the only one massless gravity. Its holographic dictionary in $D\geq5$ was established with $a$, $C_T$ and $t_2$, $t_4$ being obtained \cite{Buchel:2009sk}. The cubic gravities and even higher-order gravities are having more complexities and possibilities, extensive researches were carried out and here is the incomplete list of the references \cite{Cisterna:2018tgx,Parvizi:2017boc,Banerjee:2009fm,deBoer:2009gx,Myers:2010ru,Myers:2010jv,Brenna:2011gp,Oliva:2011xu,Dehghani:2011hm,Dehghani:2011vu,Oliva:2012zs,
Brenna:2012gp,Bazrafshan:2012rn,Dehghani:2013ldu,Ghanaatian:2014bpa,Ghanaatian:2015qra,Karasu:2016ifk,Bueno:2016dol,
Hennigar:2016gkm,Bueno:2016lrh,Chernicoff:2016qrc,Cisterna:2017umf,Hennigar:2017ego,Dykaar:2017mba,Bueno:2017sui,Ahmed:2017jod
,Bueno:2017qce,Feng:2017tev,Oliva:2010eb,Li:2017ncu,Lan:2017xcl,Ghanaatian:2018gdl,Myung:2018wya,Bueno:2018xqc,Peng:2018vbe,
Myung:2018qsn,Feng:2018qnx,Bueno:2018yzo,Wang:2018neg,Bueno:2016xff,Li:2017txk}. Particularly, in the cubic order, the natural consideration is the cubic Lovelock gravity, the relevant holographic dictionary can also be found in e.g. \cite{deBoer:2009gx,Camanho:2013pda}. However, it turns out that the class of Lovelock gravities has $t_4=0$ \cite{Buchel:2009sk,deBoer:2009gx,Camanho:2013pda}, the further investigations of other massless gravities that have non-vanishing $t_4$ are thus required. Myers quasi-topological gravity \cite{Myers:2010ru} \footnote{Specializing in $D=5$, up to a trivial six dimensional Euler density, Myers quasi-topological gravity is equivalent to the Oliva-Ray gravity \cite{Oliva:2010eb} that was constructed earlier.} is a special cubic massless gravity that admits Einstein-like black holes \cite{Myers:2010ru} and can establish $a$-theorem \cite{Myers:2010tj}. The holographic dictionary of Myers quasi-topological gravity was established in $D=5$ where $t_4$ is nontrivial \cite{Myers:2010jv}. Even more examples exist in the cubic order, for instance, Einsteinian cubic gravity was constructed in \cite{Bueno:2016xff} and its holographic dictionary was also discussed in $D=4$ \cite{Bueno:2018xqc}.

According to \cite{Camanho:2014apa}, the causality requires $(c-a)/c\ll 1$ in $d=4$, which suggests for Myers quasi-topological gravity in $D=5$ where $a\neq c$, the coupling constants should be viewed as small quantities. Subsequently, other CFT properties like the R\'enyi entropy and shear-viscosity-entropy-ratio can be expanded with the small coupling constants, and the corrections produced by the higher-order terms are expected to be controlled by $(c-a)/c$ and $t_4$, see, e.g. \cite{Banerjee:2009fm,Kats:2007mq,Hung:2011nu} (remember in other dimensions, there is $C_T$ playing the similar role as $c$-charge in $d=4$). However, both Myers quasi-topological gravity and Einsteinian gravity have only one independent coupling constant which is too strick to convince ourself that CFT parameters $a$, $C_T$ and $t_4$ can indeed control the higher-order corrections.

In this paper, we consider the generic massless cubic gravities and study their holographic dictionary in $D=5$ and $D=4$ respectively. The cubic gravities have $8$ coupling constants in total, and the ghost free condition with the decoupling of the massive scalar mode would impose two constraints, as the consequence, we are left with $6$ coupling constants to correct CFT properties. This consideration is the most generic case in the cubic order without massive modes, hence results exhibited in this paper should also apply to all cubic gravities without massive modes such as Myers quasi-topological gravity \cite{Myers:2010ru}, Einsteinian cubic gravity \cite{Bueno:2016xff} and Ricci-polynomial quasi-topological cubic gravity \cite{Li:2017ncu}. For the generic massless cubic gravities, $a$-charge and $C_T$ were already obtained in literature \cite{Li:2017txk,Li:2018drw}. In this paper, we compute $t_2$ and $t_4$ appearing in the one-point function of the energy flux (in $D=4$, there is no $t_2$ at all). Afterwards, other properties of CFT such as R\'enyi entropy and hydrodynamics should be taken into account. However, generally speaking, the nontrivial exact solutions for cubic gravities, such as black hole solutions required in the discussion of  the holographic R\'enyi entropy and hydrodynamics, are hard to come by except for some special situations \cite{Myers:2010ru,Brenna:2011gp,Brenna:2012gp,Hennigar:2016gkm,Bueno:2016lrh,Chernicoff:2016qrc,Dykaar:2017mba,Bueno:2017sui,
Feng:2017tev,Li:2017ncu,Bueno:2018xqc,Feng:2018qnx}. Fortunately, on the other hand, in general, $a\neq c$ in massless cubic gravities, which enforces us to treat the coupling constants as infinitesimal quantities in which region the black holes can be solved order by order \cite{Campanelli:1994sj,Banerjee:2009fm} \footnote{see also, e.g. \cite{Dey:2015poa,Dey:2015ytd,Mahapatra:2016dae} for more applications of the perturbative approach to black hole solutions.}. The calculations of holographic R\'enyi entropy and shear-viscosity could also be performed perturbatively \cite{Kats:2007mq,Banerjee:2009fm,Dey:2016pei,Galante:2013wta,Belin:2013dva}.

The paper is organized as follow.
\begin{itemize}
\item In section \ref{cubic-gravity}, we revisit the most generic cubic gravities in arbitrary dimensions. We present the two conditions removing massive spin-$2$ mode and massive scalar mode simultaneously and the resulting theory is the most generic massless cubic gravity for which the linearized equation of motion and the effective Newton constant $\kappa_{\rm eff}$ were reviewed. Then we present both the auxiliary-type boundary actions where auxiliary fields that should not varied in principle exist and non-auxiliary-type boundary actions without any auxiliary fields for the massless cubic gravities, including the surface term and the holographic counterterms up to the linear curvature terms. By treating the coupling constants as infinitesimal quantities, we then solve the approximate black holes with boundary topology $k$ in $D=5$ and $D=4$ respectively up to the first order.

\item In section \ref{thermodynamics}, we analyze the first order thermodynamics for the black holes obtained in section \ref{cubic-gravity}. We present the temperature and the black hole mass readily. In addition, we employ the Wald formula to obtain the black hole entropy for approximate black hole solutions. Then, we make use of two different methods to obtain the Free energy for the first order approximate black holes and verify the previous results of the mass and the entropy, during the process, we find in $D=5$ with $k\neq0$, there would exist the Casimir energy. The first law of thermodynamics is verified to be valid. Afterwards, we then analyze the thermodynamics for the second order planar black holes that are exhibited in Appendix \ref{2order}.

\item In section \ref{2-pt}, we review the results of the central charges and the holographic two-point function of the energy-momentum tensor for the massless cubic gravities. There are $a$-charge and $C_T$ coefficient playing the essential role. We also denote the notation $\tilde{a}$, $\tilde{c}$ and $\mathcal{C}_T$ that are proportional to $a$, $c$ and $C_T$ respectively for latter convenience.

\item In section \ref{flux}, we consider the one-point function of energy flux excited by a local operator which is actually a three-point function with respect to the vacuum. The excitation operator is chosen to be energy-momentum tensor with certain polarizations and the one-point function of energy flux contains two universal energy flux parameters $t_2$ and $t_4$. From the bulk point of view, we manage to obtain the energy flux parameters $t_2$ and $t_4$ for the massless cubic gravities in $D=5$. Then we obtain the parameters $\mathcal{A}$, $\mathcal{B}$ and $\mathcal{C}$ that are expected to determine the three-point function of energy-momentum tensor. Then we turn to present the physics constraints for coupling constants by requiring $C_T>0$ and non-negative energy flux in $D=5$.

\item In section \ref{entropy}, we compute the holographic R\'enyi entropy for massless cubic gravities up to the first order by using the first order approximate black holes obtained in section \ref{cubic-gravity}. In particular, we take the limit $q\rightarrow 1$, $q\rightarrow 0$ and $q\rightarrow \infty$ respectively in the R\'enyi entropy. The R\'enyi entropy with the limit $q\rightarrow 1$ recovers the entanglement entropy, and as expected, we find its behavior is proportional to $a$-charge both in $D=5$ and $D=4$. In $D=5$, we find the R\'enyi entropy with $q\rightarrow 0$ and $q\rightarrow \infty$ can be controlled by $c/a$ and $t_4$ as for more special Gauss-Bonnet gravity and Myers quasi-topological gravity. We also obtain the scaling dimension for the twist operators both in $D=5$ and $D=4$: in $D=5$, we find the scaling dimension of the twist operators is consistent with $t_2$ and $t_4$ we calculated in section \ref{flux}; in $D=4$, from the scaling dimension of the twist operators, we obtain $t_4$ parameters for massless cubic gravities in $D=4$. Subsequently, we find in deed, in $D=4$, up to the first order of the coupling constants, the R\'enyi entropy with $q\rightarrow 0$ and $q\rightarrow \infty$ can be controlled by $\mathcal{C}_T/\tilde{a}$ and $t_4$ in a variety of ways. We exhibit some examples to show that.

\item In section \ref{hydro}, we employ the ``pole method'' to calculate the shear-viscosity-entropy-ratio up to the second order associated with the second order approximate planar black holes for massless cubic gravities in $D=5$ and $D=4$ respectively. We find in $D=5$, the first order deviation from the KSS bound $1/(4\pi)$ of the shear-viscosity-entropy-ratio can be uniquely controlled by $(c-a)/c$ and $t_4$. However, we cannot find the controlling pattern in the second order only with $c$, $a$ and $t_2$, $t_4$. On the other hand, more surprisingly, we find in $D=4$, the shear-viscosity-entropy-ratio is uniquely controlled by $(\mathcal{C}_T-\tilde{a})/\mathcal{C}_T$ and $t_4$ even up to the second order.

\item In section \ref{conclusion}, the paper is summarized.

\item In Appendix \ref{2order}, we present the solutions for approximate planar black holes up to the second order of the coupling constants in $D=5$ and $D=4$.

\item In Appendix \ref{local}, we show that in the coordinates $(\rho,y)$ we adopt for calculating the one-point function of energy flux holographically, any propagators responsible for the excitation operators in $d$ dimensional CFT are localized at $\rho=\ell$ and $y^1=y^2=\cdots=y^{d-2}=0$, where $\ell$ is the effective AdS radius. The salient feature exhibited in this Appendix serves as the necessary ingredients to work out the results in section \ref{flux}.
\end{itemize}

\section{Massless cubic gravities}
\label{cubic-gravity}
\subsection{The AdS vacua and massless condition}
We consider the Einstein-gravity extended with the generic cubic curvature polynomials in $D$ dimensions coupled to a bare negative cosmological constant $\Lambda_{0}$, the bulk action is taking the following form
\be
S_{\rm bulk}=\int_{M} d^{d+1}x\sqrt{-g}L\,,\qquad L=R-2\Lambda_{0}+H^{(3)}\,,\qquad \Lambda_{0}=\fft{d(d-1)}{2\ell_0^2}\,,\label{bulk-cubic}
\ee
where $H^{(3)}$ represents the cubic polynomials and it is given by \cite{Li:2017txk}
\bea
 H^{(3)}&=& e_1 R^3 + e_2 R\,R_{\mu\nu} R^{\mu\nu} + e_3 R^{\mu}_{\nu} R^{\nu}_\rho R^{\rho}_\mu + e_4 R^{\mu\nu} R^{\rho\sigma} R_{\mu\rho\nu\sigma}\cr
&&+ e_5 R R^{\mu\nu\rho\sigma} R_{\mu\nu\rho\sigma} +e_6 R^{\mu\nu} R_{\mu \alpha\beta\gamma} R_{\nu}{}^{\alpha\beta\gamma} +
e_7 R^{\mu\nu}{}_{\rho\sigma} R^{\rho\sigma}{}_{\alpha\beta} R^{\alpha\beta}{}_{\mu\nu}\cr
&&+e_8 R^\mu{}_\nu{}^\alpha{}_\beta R^\nu{}_\rho{}^\beta{}_\gamma R^{\rho}{}_\mu{}^\gamma{}_{\alpha}\,.\label{riemanncubiclag}
\eea
It is convenient to introduce $P_{\mu\nu\rho\sigma}$ which serves as
\be
P_{\mu\nu\rho\sigma}=\fft{\partial L}{\partial R^{\mu\nu\rho\sigma}}=P^0_{\mu\nu\rho\sigma}+\sum_{i=1}^8 e_i P^{i}_{\mu\nu\rho\sigma}\,,\label{P}
\ee
where we have, explicitly \cite{Li:2018drw}
\bea
P^0_{\mu\nu\rho\sigma} &=& \ft12(g_{\mu\rho} g_{\nu\sigma} - g_{\mu\sigma} g_{\nu\rho})\,,\qquad
P^1_{\mu\nu\rho\sigma} = \ft32(g_{\mu\rho} g_{\nu\sigma} - g_{\mu\sigma} g_{\nu\rho}) R^2\,,\cr
P^2_{\mu\nu\rho\sigma} &=& \ft12  (g_{\mu\rho} g_{\nu\sigma} - g_{\mu\sigma} g_{\nu\rho}) R_{\alpha\beta}R^{\alpha\beta} \cr
&& + \ft12 R\, (g_{\mu\rho} R_{\nu\sigma} - g_{\mu\sigma} R_{\nu\rho}
-g_{\nu\rho} R_{\mu\sigma} + g_{\nu\sigma} R_{\mu\rho})\,,\cr
P^3_{\mu\nu\rho\sigma} &=& \ft34  \big(g_{\mu\rho} R_{\nu\gamma}R_{\sigma}{}^\gamma - g_{\mu\sigma} R_{\nu\gamma} R_{\rho}{}^\gamma  -
g_{\nu\rho} R_{\mu\gamma}R_{\sigma}{}^\gamma + g_{\nu\sigma} R_{\mu\gamma} R_{\rho}{}^\gamma\big)\,,\cr
P^4_{\mu\nu\rho\sigma}&=&\ft12(g_{\nu\sigma}R_{\mu\alpha\rho\beta}-g_{\nu\rho}R_{\mu\alpha\sigma\beta}-
g_{\mu\sigma}R_{\nu\alpha\rho\beta}+g_{\mu\rho}R_{\nu\alpha\sigma\beta})\cr
&&+\ft{1}{2} (R_{\mu\rho}R_{\nu\sigma}-R_{\mu\sigma}R_{\nu\rho}+R^{\alpha\beta})\,,\cr
P^5_{\mu\nu\rho\sigma}&=& \ft12(g_{\mu\rho} g_{\nu\sigma} - g_{\mu\sigma} g_{\nu\rho})R^{\alpha\beta\gamma\eta}R_{\alpha\beta\gamma\eta}+2RR_{\mu\nu\rho\sigma}\,,\cr
P^6_{\mu\nu\rho\sigma}&=&\ft{1}{2}(R_{\nu}^{\alpha}R_{\mu\alpha\rho\sigma}+
R_{\sigma}^{\alpha}R_{\mu\nu\rho\alpha}
-R_{\rho}^{\alpha}R_{\mu\nu\sigma\alpha}-R_{\mu}^{\alpha}R_{\nu\alpha\rho\sigma})\cr
&&+\ft14(g_{\nu\sigma}R_{\mu}^{\alpha\beta\gamma}-g_{\mu\sigma}R_{\nu}^{\alpha\beta\gamma})
R_{\rho\alpha\beta\gamma}
+\ft14(g_{\mu\rho}R_{\nu}^{\alpha\beta\gamma}-g_{\nu\rho}R_{\mu}^{\alpha\beta\gamma})
R_{\sigma\alpha\beta\gamma}
\,,\cr
P^7_{\mu\nu\rho\sigma}&=&3 R_{\mu\nu}^{\alpha\beta}R_{\rho\sigma\alpha\beta}\,,\qquad
P^8_{\mu\nu\rho\sigma}=\ft32  (R_{\mu}{}^{\alpha}{}_{\rho}{}^{\beta} R_{\nu\alpha\sigma\beta}-R_{\mu}{}^{\alpha}{}_{\sigma}{}^{\beta} R_{\nu\alpha\rho\beta})\,.\label{Ps}
\eea
Then, the equations of motion associated with the variation with respect to $g_{\mu\nu}$ are given by
\be
P_{\mu\rho\sigma\gamma}R_{\nu}\,^{\rho\sigma\gamma}-\fft{1}{2} g_{\mu\nu}L -2\nabla^{\rho}\nabla^{\sigma}P_{\mu\rho\sigma\nu}=0\,.\label{eom}
\ee
From the equations of motion (\ref{eom}), the cubic gravities admit AdS vacua with the effective AdS radius $\ell$
\be
ds_{\rm AdS}^2=\fft{\ell^2}{r^2}dr^2+r^2\eta_{ij}dx^idx^j\,,\label{AdS-1}
\ee
where the effective AdS radius $\ell$ is solved by equations of motion in terms of the bare AdS radius $\ell_0$ \cite{Li:2017txk}
\bea
&& \mathfrak{h}(\ell)=\fft{1}{\ell_0^2}-\fft{1}{\ell^2}- \fft{(d-5)}{(d-1)\ell^6}
\Big(d^2(d+1)^2 e_1+d^2(d+1) e_2+d^2 e_3\cr
&&+d^2 e_4+2 d(d+1) e_5+2 d e_6+4 e_7+(d-1) e_8
\Big)=0\,.
\eea
However, the excitations around the AdS vacuum (\ref{AdS-1}) have higher derivatives, and they suffer from the existence of the extra massive scalar modes and the massive spin-$2$ modes. These additional modes would cause the dual CFT non-unitary, and for our purpose, they should be removed. The decoupling of these massive modes would impose two constraints to the coupling constants associated with the cubic term $H^{(3)}$ \cite{Li:2017txk}
\bea
&&(d+1)d e_2+3d e_3+(2d-1) e_4+4(d+1)d e_5+4 (d+1) e_6+24 e_7-3 e_8=0\,,\nn\\
&&12 (d+1) d^2 e_1+\left(d^2+10d+1\right) d e_2+3 (d+1) d e_3+\left(2 d^2+5d-1\right) e_4\nn\\
&&+4 (d+5) d e_5+4 (2 d+1) e_6+3 (d-1) e_8+24 e_7=0\,.\label{ghost-free}
\eea
The cubic gravities without massive modes are referred as massless cubic gravities in \cite{Li:2018drw}. Then we have
\be
\mathfrak{h}(\ell)=\fft{1}{\ell_0^2}-\fft{1}{3\ell^6}(3\ell^4+(d-5)(d-2)(3d(d+1)e_1+2d e_2+e_4+4e_5))=0\,.\label{ell-ell0-rela}
\ee
Removing the massive modes by (\ref{ghost-free}), we are left with only graviton modes with corrections in Newton constant. Specifically, we can consider the perturbation around the AdS vacuum (\ref{AdS-1}), i.e.
\be
g_{\mu\nu}=\bar{g}_{\mu\nu}+\tilde{g}_{\mu\nu}\,,
\ee
where $\bar{g}_{\mu\nu}$ serves as the metric of the AdS background and $\tilde{g}_{\mu\nu}$ is the infinitesimal perturbation around the background. For simplicity, we impose the transverse-traceless gauge
\be
\bar{\nabla}^{\mu}\tilde{g}_{\mu\nu}=0\,,\qquad \tilde{g}_{\mu}^\mu=0\,,
\ee
where the trace is contracted by background metric $\bar{g}_{\mu\nu}$, and the covariant derivatives $\bar{\nabla}^\mu$ is also defined with respect to the AdS background (\ref{AdS-1}). Then the linearized equations of motion are given by
\be
\kappa_{\rm eff}(\bar{\Box}+\fft{2}{\ell^2})\tilde{g}_{\mu\nu}=0\,,\qquad \kappa_{\rm eff}=1+\fft{1}{\ell^4}(d-5)(d-2)(3(d+1)de_1+2de_2+e_4+4e_5)\,,\label{Newtoneff}
\ee
where the Laplacian $\bar{\Box}$ is defined with respect to the AdS vacuum (\ref{AdS-1}). Note all massless gravities have the linearized equation exactly the same as the first equation in (\ref{Newtoneff}) with different $\kappa_{\rm eff}$ for different theories, for example, for Gauss-Bonnet gravity, see e.g. \cite{Fan:2016zfs}. Recently, it was proved in \cite{Bueno:2018yzo} that the $\kappa_{\rm eff}$ can be determined by the function $\mathfrak{h}(\ell)$ appeared in (\ref{ell-ell0-rela}), it is straightforward to observe that
\be
\kappa_{\rm eff}=\fft{\ell^3}{2}\fft{\partial}{\partial \ell}\mathfrak{h}(\ell)\,.
\ee

\subsection{Boundary action}
In order to make the variation principle well-defined, we have to add the Gibbons-Hawking surface term in the action. Moreover, considering that we are evaluating the action around the AdS background where $r^2/\ell^2$ would cause the divergence, it requires the appropriate counterterms in order to obtain the finite action. Therefore, we should necessarily provide the boundary action
\be
S_{\rm tot}=S_{\rm bulk}+S_{\rm bound}\,,\qquad S_{\rm bound}=S_{\rm surf}+S_{\rm ct}\,.
\ee
However, as it was recently referred in \cite{Li:2018drw}, the massless higher order gravities suffer from the ambiguities of the surface term due to the existence of two different ways for imposing the well-defined variation associated with the metric $g_{\mu\nu}$. Consequently, one can have two sets of boundary actions leading to exactly same holographic results like holographic one-point functions of energy-momentum tensor \cite{Li:2018drw}. The first way is to introduce the so called auxiliary field $\Phi^\mu_\nu$ in the surface term that should not be involved in the variation of the metric $g_{\mu\nu}$ \cite{Deruelle:2009zk}
\be
S_{\rm surf-aux}=\fft{1}{4\pi}\int_{\partial M} d^dx\sqrt{-h} \Phi^\mu_\nu K_{\mu}^{\nu}\,,\qquad
\Phi^\mu_\nu=P^\mu{}_{\rho\nu\sigma} n^\rho n^\sigma\,.\label{surf}
\ee
Note $P^\mu\,_{\rho\nu\sigma}$ can be found in (\ref{P}) and (\ref{Ps}) \cite{Li:2018drw}, and $K_\mu^\nu$ is the extrinsic curvature which is defined as
\be
K_{\mu\nu}=h^\rho_\mu \nabla_\rho n_\nu\,,
\ee
where $n_\mu$ is given by
\be
n^\mu=\fft{r}{\ell}(\fft{\partial}{\partial r})^\mu\,.
\ee
The resulting boundary action is referred as the auxiliary-type in this paper, and the auxiliary type boundary action applies to the generic higher order gravities, see, e.g. \cite{Deruelle:2009zk,Liu:2017kml,Li:2017ncu}.

In massless gravities, it is remarkable that the auxiliary field is not necessary in the construction of the surface term, and the variation with respect to $\delta g_{\mu\nu}$ demands varying all fields involved in the action.
\be
S_{\rm surf-naux}=\fft{1}{4\pi}\int_{\partial M} d^dx\sqrt{-h}\, \widetilde P_{\mu\nu\rho\sigma}\, K^{\mu\rho} \, n^\nu n^\sigma\,,\label{surf1}
\ee
where we introduced
\be
\widetilde P_{\mu\nu\rho\sigma} = P^0_{\mu\nu\rho\sigma} + \fft15\sum_{i=1}^8 e_i P^i_{\mu\nu\rho\sigma}\,.
\ee
We refer the resulting boundary action as the non-auxiliary-type boundary action. The boundary action of Gauss-Bonnet gravity and more general Lovelock gravity \cite{Myers:1987yn,Liu:2008zf,Liu:2017kml} is of this type. Typically, in Gauss-Bonnet gravity and Lovelock gravity, the surface term (\ref{surf1}) can be expressed in terms of the extrinsic curvature $K_{\mu\nu}$ \cite{Myers:1987yn,Liu:2008zf,Liu:2017kml}.

We now provide the explicit holographic counterterms in both auxiliary-type boundary action and non-auxiliary-type boundary action of the massless cubic gravities up to the linear curvature term.
The auxiliary-type counterterms take as follows
\bea
S_{\rm ct-aux} &=& -2\int d^dx\sqrt{-h}\Big(\big(\fft{d-1}{\ell}+\fft{(d-2)(d-1)^2}{\ell^5}\big(3d(d+1)e_1+2de_2+e_4+e_5))
\cr &&
\cr && +\big(\fft{\ell}{2(d-2)}+\fft{d-1}{2\ell^3}(3d(d+1)e_1+2d e_2+e_4+4e_5)\big)\mathcal{R}\Big)+\cdots\,,\label{ct-aux}
\eea
where $\mathcal{R}$ is the ${\rm Ricci}$ scalar curvature associated with the boundary metric. The non-auxiliary-type counterterms are given by
\bea
S_{\rm ct-naux} &=& -2\int d^dx\sqrt{-h}\Big(\big(\fft{d-1}{\ell}+\fft{(d-5)(d-2)(d-1)}{5\ell^5}\big(3d(d+1)e_1+2de_2+e_4+e_5))
\cr &&
\cr && +\big(\fft{\ell}{2(d-2)}+\fft{d+3}{10\ell^3}(3d(d+1)e_1+2d e_2+e_4+4e_5)\big)\mathcal{R}\Big)+\cdots\,.\label{ct-noaux}
\eea
Throughout this paper, we mainly consider the holographic aspects of the massless cubic gravities in $D=5$, $d=4$ and $D=4$, $d=3$. For our purpose, the linear curvature counterterms are adequate. In higher dimensions, one has to add more counterterms that are given by higher order of boundary curvature invariants to cancel the divergence. For Einstein gravity, they can be found in \cite{Balasubramanian:1999re,deHaro:2000vlm}; for Gauss-Bonnet gravity, they can be found in \cite{Liu:2008zf}.

\subsection{Approximate solutions of the black holes}
In this subsection, we intend to obtain the black holes of the generic massless cubic gravities. Unfortunately, it turns out that the exact black hole solutions can only exist provided with some further constraints of the coupling constants associated with the cubic terms, like quasi-topological gravities \cite{Myers:2010ru,Li:2017ncu} and Einsteinian cubic gravity \cite{Hennigar:2016gkm,Bueno:2016lrh,Bueno:2018xqc}, while for the generic coupling constants, there is no exact solution. In this paper, instead of trying to find the exact solutions, we treat coupling constants of cubic terms $e_i$ as the infinitesimal quantities compared to any other relevant quantities in theory (for example, $\ell_0$), and we can then perform the perturbative method proposed in \cite{Campanelli:1994sj} to solve out the black holes order by order. This treatment is consistent with the causality \cite{Camanho:2014apa}, and we will see it is also consistent with other physical constraints in section \ref{flux} and section \ref{entropy}. In this subsection, we follow the procedure in \cite{Campanelli:1994sj} and present the approximate solutions of the black holes up to the first order of $e_i$ for the generic massless cubic gravities in $D=5$ and $D=4$. In Appendix \ref{2order}, we present the approximate black hole solutions with the flat boundary topology up to the second order of $e_i$ for the purpose of computing the shear-viscosity in section \ref{hydro}.

The metric ansatz is given by
\be
ds^2=-f(r)dt^2+\fft{1}{h(r)}dr^2+r^2 d\Omega_{k,D-2}^2\,,\label{bh}
\ee
where $k=-1,0,1$ is referred as the $(D-2)$-hyperbolic space, $(D-2)$-flat space and the $(D-2)$-sphere respectively. In the construction of the approximate black holes order by order, each order will include one integration constant. However, it should be emphasized that the infinitesimal correction produced by higher order curvature invariants shall not alter the horizon $r_0$ of the uncorrected black holes, otherwise the perturbative approach with infinitesimal coupling constants $e_i$ would break down in the sense that all $\mathcal{O}(e_i^{n})$ terms should share the same magnitude order to ensure $f(r_h)=h(r_h)=0$ at the new horizon $r_h$. In other words, for black holes solved by the perturbative approach, each order itself should vanish at the horizon $r_0$, leaving the black hole solutions with only one parameter: the horizon or, equivalently, the mass. This result is consistent with the no-hair theorem and it also reflects the fact that the spectrum contains sole massless graviton. In this paper, we follow the prescription mentioned above to construct the black holes perturbatively.

In $D=5$, we have
\bea
f(r) &=& \frac{1}{3 \ell_0^6 r^{10} r_0^2}(3 k r_0^2 r^{10}+(2 k^3 (1200 e_1+340 e_2+108 e_3+83 e_4+328 e_5+76 e_6)-3 k r_0^4) r^8
\cr && -72 k^3 (90 e_1+22 e_2+6 e_3+5 e_4+18 e_5+4 e_6) r_0^6 r^2+2 k^3 (2040 e_1+452 e_2+108 e_3
\cr &&+97 e_4+320 e_5+68 e_6) r_0^8) \ell_0^6+3 r_0^2 (r^2-r_0^2) (r^{10}+r_0^2 r^8+4 k^2 (930 e_1+254 e_2
\cr &&+78 e_3+61 e_4+234 e_5+54 e_6) r^6+4 k^2 (930 e_1+254 e_2+78 e_3+61 e_4+234 e_5+54 e_6) r_0^2 r^4
\cr &&+2 k^2 (120 e_1+76 e_2+36 e_3+23 e_4+112 e_5+28 e_6) r_0^4 r^2-2 k^2 (2040 e_1+452 e_2+108 e_3
\cr &&+97 e_4+320 e_5+68 e_6) r_0^6) \ell_0^4+6 k r_0^4 (204 e_3 r^8+161 e_4 r^8+608 e_5 r^8+140 e_6 r^8
\cr &&-240 e_3 r_0^4 r^4-198 e_4 r_0^4 r^4-712 e_5 r_0^4 r^4-160 e_6 r_0^4 r^4-72 e_3 r_0^6 r^2-60 e_4 r_0^6 r^2
\cr &&-216 e_5 r_0^6 r^2-48 e_6 r_0^6 r^2+108 e_3 r_0^8+97 e_4 r_0^8+320 e_5 r_0^8+68 e_6 r_0^8
\cr &&+120 e_1 (21 r^8-29 r_0^4 r^4-9 r_0^6 r^2+17 r_0^8)+4 e_2 (169 r^8-216 r_0^4 r^4-66 r_0^6 r^2+113 r_0^8)) \ell_0^2
\cr &&+2 r_0^2 (r^4-r_0^4) (e_4 r^8+4 e_5 r^8+252 e_3 r_0^4 r^4+200 e_4 r_0^4 r^4+748 e_5 r_0^4 r^4+172 e_6 r_0^4 r^4
\cr &&-108 e_3 r_0^8-97 e_4 r_0^8-320 e_5 r_0^8-68 e_6 r_0^8+60 e_1 (r^8+53 r_0^4 r^4-34 r_0^8)
\cr &&+e_2 (8 r^8+844 r_0^4 r^4-452 r_0^8))+\mathcal{O}(e_i)^2\,,\label{f-k-D=5}
\eea
and
\bea
h(r) &=& \frac{1}{3 \ell_0^6 r^{10} r_0^2}(3 k r_0^2 r^{10}+(2 k^3 (1200 e_1+340 e_2+108 e_3+83 e_4+328 e_5+76 e_6)-3 k r_0^4) r^8
\cr &&-48 k^3 (240 e_1+72 e_2+24 e_3+18 e_4+72 e_5+17 e_6) r_0^6 r^2+2 k^3 (4560 e_1+1388 e_2
\cr &&+468 e_3+349 e_4+1400 e_5+332 e_6) r_0^8) \ell_0^6+3 r_0^2 (r^{12}+(4 k^2 (930 e_1+254 e_2+78 e_3
\cr &&+61 e_4+234 e_5+54 e_6)-r_0^4) r^8-2 k^2 (2580 e_1+744 e_2+240 e_3+183 e_4+716 e_5
\cr &&+168 e_6) r_0^4 r^4-32 k^2 (240 e_1+72 e_2+24 e_3+18 e_4+72 e_5+17 e_6) r_0^6 r^2
\cr &&+2 k^2 (4560 e_1+1388 e_2+468 e_3+349 e_4+1400 e_5+332 e_6) r_0^8) \ell_0^4
\cr &&+6 k r_0^4 (204 e_3 r^8+161 e_4 r^8+608 e_5 r^8+140 e_6 r^8-480 e_3 r_0^4 r^4-366 e_4 r_0^4 r^4
\cr &&-1432 e_5 r_0^4 r^4-336 e_6 r_0^4 r^4-192 e_3 r_0^6 r^2-144 e_4 r_0^6 r^2-576 e_5 r_0^6 r^2-136 e_6 r_0^6 r^2
\cr &&+468 e_3 r_0^8+349 e_4 r_0^8+1400 e_5 r_0^8+332 e_6 r_0^8+120 e_1 (21 r^8-43 r_0^4 r^4-16 r_0^6 r^2+38 r_0^8)
\cr &&+4 e_2 (169 r^8-372 r_0^4 r^4-144 r_0^6 r^2+347 r_0^8)) \ell_0^2+2 r_0^2 (r^4-r_0^4) (e_4 r^8+4 e_5 r^8
\cr &&+252 e_3 r_0^4 r^4+200 e_4 r_0^4 r^4+748 e_5 r_0^4 r^4+172 e_6 r_0^4 r^4-468 e_3 r_0^8-349 e_4 r_0^8
\cr &&-1400 e_5 r_0^8-332 e_6 r_0^8+60 e_1 (r^8+53 r_0^4 r^4-76 r_0^8)
\cr &&+4 e_2 (2 r^8+211 r_0^4 r^4-347 r_0^8))+\mathcal{O}(e_i^2)\,,\label{h-k-D=5}
\eea
where $r_0$ is the radius that the event horizon is located. It is of interests to have a close look at the effective AdS radius in this approximate solutions (\ref{f-k-D=5}) and (\ref{h-k-D=5}). The effective radius is encoded in the coefficient of $r^2$ in solutions, and we have
\be
\fft{1}{\ell^2}=\fft{1}{\ell_0^2}+\fft{2(60e_1+8e_2+e_4+4e_5)}{3\ell_0^6}+\mathcal{O}(e_i^2)\,.\label{1-ord-ell-5D}
\ee
In Appendix \ref{2order}, we can see the effective radius up to the second order is given by
\be
\fft{1}{\ell^2}=\fft{1}{\ell_0^2}+\fft{2(60e_1+8e_2+e_4+4e_5)}{3\ell_0^6}+\fft{4(60e_1+8e_2+e_4+4e_5)^2}{3\ell_0^{10}}+\mathcal{O}(e_i)^3\,.
\label{2-ord-ell-5D}
\ee
It can be easily verified that (\ref{1-ord-ell-5D}) and (\ref{2-ord-ell-5D}) is consistent with (\ref{ell-ell0-rela}) up to $e_i$ and $e_i^2$ correspondingly.

In $D=4$, we have
\bea
f(r) &=& \frac{1}{6 \ell_0^6 r^7 r_0^3}(6 k r_0^3 r^7+(k^3 (252 e_1+78 e_2+27 e_3+22 e_4+52 e_5+18 e_6)-6 k r_0^4) r^6
\cr &&-27 k^3 (48 e_1+12 e_2+3 e_3+3 e_4+8 e_5+2 e_6) r_0^5 r+k^3 (1044 e_1+246 e_2+54 e_3
\cr &&+59 e_4+164 e_5+36 e_6) r_0^6) \ell_0^6+3 r_0^2 (2 r_0 r^9+(9 k^2 (60 e_1+18 e_2+6 e_3+5 e_4
\cr &&+12 e_5+4 e_6)-2 r_0^4) r^6-2 k^2 (360 e_1+96 e_2+27 e_3+25 e_4+64 e_5+18 e_6) r_0^3 r^3
\cr &&-18 k^2 (48 e_1+12 e_2+3 e_3+3 e_4+8 e_5+2 e_6) r_0^5 r+k^2 (1044 e_1+246 e_2+54 e_3
\cr &&+59 e_4+164 e_5+36 e_6) r_0^6) \ell_0^4+3 k r_0^4 (81 e_3 r^6+68 e_4 r^6+164 e_5 r^6+54 e_6 r^6
\cr &&-108 e_3 r_0^3 r^3-100 e_4 r_0^3 r^3-256 e_5 r_0^3 r^3-72 e_6 r_0^3 r^3-27 e_3 r_0^5 r-27 e_4 r_0^5 r-72 e_5 r_0^5 r
\cr &&-18 e_6 r_0^5 r+54 e_3 r_0^6+59 e_4 r_0^6+164 e_5 r_0^6+36 e_6 r_0^6+36 e_1 (23 r^6-40 r_0^3 r^3-12 r_0^5 r+29 r_0^6)
\cr &&+6 e_2 (41 r^6-64 r_0^3 r^3-18 r_0^5 r+41 r_0^6)) \ell_0^2+r_0^3 (r^3-r_0^3) (4 e_4 r^6+16 e_5 r^6+108 e_3 r_0^3 r^3
\cr &&+91 e_4 r_0^3 r^3+220 e_5 r_0^3 r^3+72 e_6 r_0^3 r^3-54 e_3 r_0^6-59 e_4 r_0^6-164 e_5 r_0^6-36 e_6 r_0^6
\cr &&+36 e_1 (4 r^6+31 r_0^3 r^3-29 r_0^6)+6 e_2 (4 r^6+55 r_0^3 r^3-41 r_0^6))+\mathcal{O}(e_i^2)\,,\label{f-k-D=4}
\eea
and
\bea
h(r) &=& \frac{1}{6 \ell_0^6 r^7 r_0^3}(6 k r_0^3 r^7+(k^3 (252 e_1+78 e_2+27 e_3+22 e_4+52 e_5+18 e_6)-6 k r_0^4) r^6
\cr &&-27 k^3 (72 e_1+24 e_2+9 e_3+7 e_4+16 e_5+6 e_6) r_0^5 r+k^3 (1692 e_1+570 e_2
\cr &&+216 e_3+167 e_4+380 e_5+144 e_6) r_0^6) \ell_0^6+3 r_0^2 (2 r_0 r^9+(9 k^2 (60 e_1+18 e_2
\cr &&+6 e_3+5 e_4+12 e_5+4 e_6)-2 r_0^4) r^6-2 k^2 (468 e_1+150 e_2+54 e_3+43 e_4
\cr &&+100 e_5+36 e_6) r_0^3 r^3-18 k^2 (72 e_1+24 e_2+9 e_3+7 e_4+16 e_5+6 e_6) r_0^5 r
\cr &&+k^2 (1692 e_1+570 e_2+216 e_3+167 e_4+380 e_5+144 e_6) r_0^6) \ell_0^4
\cr &&+3 k r_0^4 (81 e_3 r^6+68 e_4 r^6+164 e_5 r^6+54 e_6 r^6-216 e_3 r_0^3 r^3-172 e_4 r_0^3 r^3
\cr &&-400 e_5 r_0^3 r^3-144 e_6 r_0^3 r^3-81 e_3 r_0^5 r-63 e_4 r_0^5 r-144 e_5 r_0^5 r-54 e_6 r_0^5 r
\cr &&+216 e_3 r_0^6+167 e_4 r_0^6+380 e_5 r_0^6+144 e_6 r_0^6+36 e_1 (23 r^6-52 r_0^3 r^3-18 r_0^5 r
\cr &&+47 r_0^6)+6 e_2 (41 r^6-100 r_0^3 r^3-36 r_0^5 r+95 r_0^6)) \ell_0^2
\cr &&+r_0^3 (r^3-r_0^3) (4 e_4 r^6+16 e_5 r^6+108 e_3 r_0^3 r^3+91 e_4 r_0^3 r^3+220 e_5 r_0^3 r^3
\cr &&+72 e_6 r_0^3 r^3-216 e_3 r_0^6-167 e_4 r_0^6-380 e_5 r_0^6-144 e_6 r_0^6
\cr &&+36 e_1 (4 r^6+31 r_0^3 r^3-47 r_0^6)+6 e_2 (4 r^6+55 r_0^3 r^3-95 r_0^6))+\mathcal{O}(e_i^2)\,.\label{h-k-D=4}
\eea
Up to the first order, the effective AdS radius is
\be
\fft{1}{\ell^2}=\fft{1}{\ell_0^2}+\fft{2(36e_1+6e_2+e_4+4e_5)}{3\ell_0^6}+\mathcal{O}(e_i^2)\,,\label{1-ord-ell-4D}
\ee
which is consistent with (\ref{ell-ell0-rela}) up to $e_i$. According to the result in Appendix, the effective AdS radius up to the second order is given by
\be
\fft{1}{\ell^2}=\fft{1}{\ell_0^2}+\fft{2(36e_1+6e_2+e_4+4e_5)}{3\ell_0^6}+\fft{4(36e_1+6e_2+e_4+4e_5)^2}{3\ell_0^{10}}+\mathcal{O}(e_i)^3\,,
\label{2-ord-ell-4D}
\ee
which is consistent with (\ref{ell-ell0-rela}) up to $e_i^2$. We shall discuss the thermodynamics of the approximate black holes in the next section.

\section{Thermodynamics}
\label{thermodynamics}
In this section, we present the black hole thermodynamics for the approximate black holes obtained in the previous section, i.e. the first order thermodynamics in $D=5$ and $D=4$. Additionally, we also exhibit the results of the second order planar black holes, i.e. $k=0$, in $D=5$ and $D=4$, in which the approximate solutions are given in Appendix \ref{2order}.
\subsection{The first order in $D=5$}
The temperature of a black hole is given by
\be
T=\fft{1}{4\pi}\sqrt{f'(r_0)h'(r_{0})}\,,\label{T-form}
\ee
where the primes stands for the derivative over $r$. For black holes with (\ref{f-k-D=5}) and (\ref{h-k-D=5}) in $D=5$, we have explicitly
\bea
T&=&\frac{1}{6 \pi \ell_0^6 r_0^5}(\ell_0^6 (4 (150 e_1+50 e_2+18 e_3+13 e_4+62 e_5+14 e_6) k^3+3 k r_0^4)
\cr && +6 \ell_0^4 (2 (60 e_1+8 e_2+e_4+4 e_5) k^2 r_0^2+r_0^6)-12 (30 e_1+34 e_2+18 e_3
\cr &&+11 e_4+54 e_5+14 e_6) \ell_0^2 k r_0^4-8 (30 e_1+34 e_2+18 e_3+11 e_4
\cr && +54 e_5+14 e_6) r_0^6)+\mathcal{O}(e_i^2)\,.\label{T-5D-1order}
\eea
To compute the black hole entropy, we take use of the Wald formula \cite{Wald:1993nt,Iyer:1994ys}
\be
S=\Big(-2\pi\int d^{D-2}x\sqrt{\sigma}P_{\mu\nu\rho\sigma}\varepsilon^{\mu\nu}\varepsilon^{\rho\sigma}\Big)_{r=r_0}\,,\label{Wald1}
\ee
where $\sigma$ is the induced metric in the space-like $(D-2)$-boundary, and $\varepsilon_{\mu\nu}$ is the binormal to the horizon. By binormal, we mean, explicitly
\be
\varepsilon=\sqrt{\fft{h}{f}}\,dt\wedge dr\,,\qquad \varepsilon_{\mu\nu}\varepsilon^{\mu\nu}=-2\,.
\ee
For static black holes where $P_{\mu\nu\rho\sigma}\varepsilon^{\mu\nu}\varepsilon^{\rho\sigma}$ is constant on horizon, we shall have a simpler formula taking the form as
\be
S=-2\pi\,\omega_{k,D-2}\,r_0^{D-2}\,\Big(P_{\mu\nu\rho\sigma}\varepsilon^{\mu\nu}\varepsilon^{\rho\sigma}\Big)_{r=r_0}\,,\label{Wald2}
\ee
where $\omega_{k,D-2}$ is the volume of a unit $(D-2)$-boundary. To be precise, for $k=1$ it is the volume of a unit $S^{D-2}$, while for $k=-1,0$, this ``volume'' might be infinite, so it is natural to divide this factor out and the corresponding $S$ is viewed as the entropy density $s$. Throughout this paper, we always keep the factor $\omega_{k,D-2}$, but the convention is settled such that $\omega_{k,D-2}$ for $k=1$ is the finite volume of a unit $(D-2)$-sphere, while $\omega_{k,D-2}$ for $k=-1,0$ is set to $1$ and the corresponding thermodynamic quantity is actually the density, for example, the mass density denoted as $m$, the entropy density denoted as $s$ and the Free energy density denoted as $\mathcal{F}$. Explicitly, for $D=5$, we have
\bea
S &=& -2 \pi  r_0^3 \Big(\frac{48 (90 e_1+22 e_2+6 e_3+5 e_4+18 e_5+4 e_6)}{\ell_0^4}
\cr &&+\frac{96 (90 e_1+22 e_2+6 e_3+5 e_4+18 e_5+4 e_6) k}{\ell_0^2 r_0^2}
\cr &&+\frac{12 (300 e_1+80 e_2+24 e_3+19 e_4+68 e_5+16 e_6) k^2}{r_0^4}-2\Big)\omega_{k,3}+\mathcal{O}(e_i^2)\,.\label{entropy-5D-1order}
\eea
We then turn to obtain the mass of the black hole. The mass can be read off from the asymptotic expansion of $f(r)$, i.e.
\be
f(r)=\fft{r^2}{\ell^2}+k+\fft{f^{(d)}}{r^{d-2}}+\cdots\,,
\ee
where $\ell$ is the effective AdS radius and the coefficient $f^{(2)}$ is the mass parameter. However, we have a modified Newton constant $\kappa_{\rm eff}$ now, and the mass is given by the following formula
\be
M=-(D-3)\kappa_{\rm eff}\int d^{D-2}\sqrt{\sigma}f^{(d)}\,,\label{Mass}
\ee
where $\kappa_{\rm eff}$ is given in (\ref{Newtoneff}) with $\ell$ the effective AdS radius. The formula (\ref{Mass}) can be verified to be true by using the holographic energy-momentum tensor formula in \cite{Li:2018drw}. Even though in \cite{Li:2018drw}, the formula was derived under the flat boundary assumption, the results apply to the general curved background, because additional terms due to the curved background are divergent and can be canceled out identically by using the counterterms (\ref{ct-aux}) or (\ref{ct-noaux}). Substitute (\ref{1-ord-ell-5D}) into (\ref{Newtoneff}), we immediately have
\be
\kappa_{\rm eff}=1-\fft{2(60e_1+8e_2+e_4+4e_5)}{\ell_0^4}+\mathcal{O}(e_i^2)\,.
\ee
Then the mass formula (\ref{Mass}) yields
\bea
M &=& \frac{\omega_{k,3}}{\ell_0^6 r_0^2}(\ell_0^6 (3 k r_0^4-2 (1200 e_1+340 e_2+108 e_3+83 e_4+328 e_5+76 e_6) k^3)
\cr && +3 \ell_0^4 (r_0^6-4 (930 e_1+254 e_2+78 e_3+61 e_4+234 e_5+54 e_6) k^2 r_0^2)
\cr &&-12 (1290 e_1+342 e_2+102 e_3+81 e_4+306 e_5+70 e_6) \ell_0^2 k r_0^4
\cr &&-4 (1650 e_1+430 e_2+126 e_3+101 e_4+378 e_5+86 e_6) r_0^6)+\mathcal{O}(e_i^2)\,.\label{mass-5D-1order}
\eea
It is then can be readily verified from (\ref{T-5D-1order}), (\ref{entropy-5D-1order}) and (\ref{mass-5D-1order}) that the first law of thermodynamics is valid
\be
dM=TdS\,.\label{1slaw}
\ee

The Free energy plays an essential role in determining the thermodynamics, we shall consider the Free energy of the black holes and verify our results of the black hole mass and entropy. For black holes, the Free energy can be derived by evaluating the on-shell Euclidean action where the time direction has been Wick rotated, i.e. $t\rightarrow -{\rm i}\tau$. In Euclidean action, the partition function is determined by the Free energy according to the formula
\be
Z=e^{S^{\rm Euc}}(T)=e^{-\fft{F}{T}}\,,\qquad S_{\rm bulk}^{\rm Euc}(T)=\int_0^{\fft{1}{T}} d\tau \int_{r_0}^{\fft{1}{\epsilon}} dr \int d^3x\sqrt{\sigma}L\,,\label{F-partition}
\ee
where $\epsilon\rightarrow 0$ is the UV cutoff. Therefore we have
\be
F=-T S_{\rm bulk}^{\rm Euc}(T)\,.\label{F1}
\ee
However, as it is mentioned in section \ref{cubic-gravity}, the AdS background would cause the UV divergence of the on-shell action. There are two approaches to fix the divergence. One approach is to subtract the AdS background contribution which is given by
\be
S_{\rm bulk}^{(0) \rm Euc}(\tilde{T})=\int_0^{\fft{1}{\tilde{T}}} d\tau \int_{r_0}^{\fft{1}{\epsilon}} dr \int d^3x\sqrt{\sigma}L\,,
\ee
where $\tilde{T}$ is different from $T$ due to the red-shift effect, and it is given by
\be
\fft{T}{\tilde{T}}=\fft{f(r)}{r^2/\ell^2}\Big|_{r=1/\epsilon}\,,
\ee
in which $\ell$ is the effective AdS radius. We then have
\be
F=-T(S_{\rm bulk}^{ \rm Euc}(T)-S_{\rm bulk}^{(0) \rm Euc}(\tilde{T}))\,.\label{F2}
\ee
Notice for black holes (\ref{f-k-D=5}) and (\ref{h-k-D=5}), the approximate AdS background is
\be
f^{(0)}(r)=h^{(0)}(r)=\fft{r^2}{\ell^2}+k\,,\label{vac-5D-order-1}
\ee
where $\ell$ takes the value in (\ref{1-ord-ell-5D}). Substituting (\ref{f-k-D=5}), (\ref{h-k-D=5}) and (\ref{vac-5D-order-1}) into (\ref{F2}) yields
\bea
F &=& \frac{\omega_{k,3}}{3 \ell_0^6 r_0^2}(\ell_0^6 (2 (1200 e_1+220 e_2+36 e_3+41 e_4-8 e_5+4 e_6) k^3+3 k r_0^4)
\cr && +\ell_0^4 (12 (1050 e_1+230 e_2+54 e_3+49 e_4+130 e_5+30 e_6) k^2 r_0^2-3 r_0^6)
\cr && +12 (1590 e_1+362 e_2+90 e_3+79 e_4+270 e_5+58 e_6) \ell_0^2 k r_0^4+4 (1650 e_1\cr &&+430 e_2
+126 e_3+101 e_4+378 e_5+86 e_6) r_0^6)+\mathcal{O}(e_i^2)\,.\label{Free-5D-order-1}
\eea
One can immediately verify that up to the first order, we have
\be
F=M-TS\,,\qquad M=-T^2\fft{\partial}{\partial T}\Big(\fft{F}{T}\Big)\,,\qquad S=-\fft{\partial F}{\partial T}\,.\label{ther-rela}
\ee

There is another approach to obtain the finite Free energy. We can calculate the total action involving the auxiliary-type boundary action (\ref{surf}) and (\ref{ct-aux}), or the non-auxiliary-type boundary action (\ref{surf1}) and (\ref{ct-noaux}) instead of only the bulk action. However, this approach suffers from one subtlety for black holes with curved boundary in odd bulk dimension. In odd bulk dimension, there would be the additional Casimir energy for black holes with curved boundary \cite{Lu:2014maa}
\be
\tilde{F}=-T S^{\rm Euc}_{\rm tot}(T)=F+M_{\rm Casi}\,.\label{F3}
\ee
Subtracting the Casimir energy would give rise to the correct Free energy. The Casimir energy can be obtained by evaluating the Free energy of AdS vacua
\be
M_{\rm Casi}=-T_{\rm arb} \,S^{(0)\rm Euc}_{\rm tot}(T_{\rm arb})\,,\label{Casi}
\ee
where $T_{\rm arb}$ is an arbitrary temperature. For black holes (\ref{f-k-D=5}) and (\ref{h-k-D=5}) in $D=5$, the Casimir energy is obtained by using (\ref{vac-5D-order-1}) and (\ref{1-ord-ell-5D})
\be
M_{\rm Casi}=-\fft{k^2}{4\ell_0^2}\Big(3\ell_0^4+16(60e_1+8e_2+e_4+4e_5)\Big)+\mathcal{O}(e_i^2)\,.\label{Casi-D=5}
\ee
It is evidently that for flat boundary the Casimir energy vanishes identically. Evaluating the total action (auxiliary-type or non-auxiliary type) and subtracting the Casimir energy (\ref{Casi-D=5}) reproduces (\ref{Free-5D-order-1}).
\subsection{The first order in $D=4$}
For black holes with (\ref{f-k-D=4}) and (\ref{f-k-D=4}), by applying (\ref{T-form}), the temperature is given by
\bea
T&=&\frac{1}{8 \pi  \ell_0^6 r_0^5}(\ell_0^6 (2 k r_0^4-(36 e_1+6 e_2+e_4+4 e_5) k^3)
\cr && +3 \ell_0^4 (2 r_0^6
-(108 e_1+30 e_2+9 e_3+8 e_4+20 e_5+6 e_6) k^2 r_0^2)
\cr && -9 (60 e_1+18 e_2+6 e_3
+5 e_4+12 e_5+4 e_6) \ell_0^2 k r_0^4
\cr && -9 (12 e_1+6 e_2+3 e_3+2 e_4+4 e_5+2 e_6) r_0^6)+\mathcal{O}(e_i^2)\,.\label{T-4D-1order}
\eea
Then applying (\ref{Wald2}) exactly like we do previously, we have the corresponding entropy
\bea
S &=& -2 \pi  r_0^2 \Big(\frac{9 (48 e_1+12 e_2+3 e_3+3 e_4+8 e_5+2 e_6)}{\ell_0^4}
\cr && +\frac{18 (48 e_1+12 e_2+3 e_3+3 e_4+8 e_5+2 e_6) k}{\ell_0^2 r_0^2}
\cr &&+\frac{(288 e_1+84 e_2+27 e_3+23 e_4+56 e_5+18 e_6) k^2}{r_0^4}-2\Big)\omega_{k,2}+\mathcal{O}(e_i^2)\,.\label{entropy-4D-1order}
\eea
Furthermore, in $D=4$, we have the effective Newton constant as follows
\be
\kappa_{\rm eff}=1-\fft{2(36e_1+6e_2+e_4+4e_5)}{\ell_0^4}+\mathcal{O}(e_i^2)\,,
\ee
subsequently, the mass formula (\ref{mass-5D-1order}) gives rise to
\bea
M &=&\Big(-\frac{3 (156 e_1+42 e_2+12 e_3+11 e_4+28 e_5+8 e_6) r_0^3}{\ell_0^6}
\cr && -\frac{9 (108 e_1+30 e_2+9 e_3+8 e_4+20 e_5+6 e_6) k r_0}{\ell_0^4}
\cr &&+\frac{2 r_0^4-9 (60 e_1+18 e_2+6 e_3+5 e_4+12 e_5+4 e_6) k^2}{\ell_0^2 r_0}
\cr &&-\frac{(252 e_1+78 e_2+27 e_3+22 e_4+52 e_5+18 e_6) k^3}{3 r_0^3}+2 k r_0\Big)\omega_{k,2}+\mathcal{O}(e_i^2)\,.
\eea
We can verify (\ref{1slaw}).

Given the vacuum (\ref{vac-5D-order-1}) with (\ref{1-ord-ell-4D}) and the black holes (\ref{f-k-D=4}) and (\ref{h-k-D=4}), either using (\ref{F2}) or (\ref{F3}) (there is no Casimir energy in $D=4$), one has the same results of the Free energy
\bea
F&=&\frac{\omega_{k,2}}{6 \ell_0^6 r_0^3}(\ell_0^6 ((468 e_1+114 e_2+27 e_3+28 e_4+76 e_5+18 e_6) k^3+6 k r_0^4)
\cr && +\ell_0^4 (9 (324 e_1+78 e_2+18 e_3+19 e_4+52 e_5+12 e_6) k^2 r_0^2-6 r_0^6)
\cr &&+27 (180 e_1+42 e_2+9 e_3+10 e_4+28 e_5+6 e_6) \ell_0^2 k r_0^4+9 (156 e_1+42 e_2
\cr && +12 e_3+11 e_4+28 e_5+8 e_6) r_0^6)+\mathcal{O}(e_i^2)\,.
\eea
It follows we have (\ref{ther-rela}).
\subsection{The second order results with $k=0$}
In this subsection, we present the thermodynamical quantities associated with the second order planar black holes ($k=0$) provided in Appendix \ref{2order}.

In $D=5$, the solutions are given in (\ref{f-k=0-D=5-2order}) and (\ref{h-k=0-D=5-2order}), we have the temperature
\bea
T &=& \frac{r_0}{3 \pi  \ell_0^{10}} (-72 e_3 \ell_0^4-44 e_4 \ell_0^4-216 e_5 \ell_0^4-56 e_6 \ell_0^4+8 e_2 (114624 e_3+85193 e_4
\cr && +360612 e_5+77336 e_6-17 \ell_0^4)-120 e_1 (-90212 e_2-28092 e_3-20659 e_4
\cr &&-88856 e_5-18948 e_6+\ell_0^4)+19526400 e_1^2+1485856 e_2^2+140256 e_3^2
\cr &&+78244 e_4^2+1371456 e_5^2+63904 e_6^2+210600 e_3 e_4+877920 e_3 e_5+662184 e_4 e_5
\cr &&+189312 e_3 e_6+142184 e_4 e_6+592160 e_5 e_6+3 \ell_0^8)+\mathcal{O}(e_i^3)\,.\label{T-order2-D=5}
\eea
The entropy density is
\bea
s &=& \frac{4 \pi  r_0^3}{\ell_0^8} (-24 (90 e_1+22 e_2+6 e_3+5 e_4+18 e_5+4 e_6) \ell_0^4+128 (159300 e_1^2
\cr && +60 (1498 e_2+474 e_3+347 e_4+1494 e_5+322 e_6) e_1+12452 e_2^2+1188 e_3^2
\cr &&+665 e_4^2+11556 e_5^2+544 e_6^2+1788 e_3 e_4+7416 e_3 e_5+5604 e_4 e_5+1608 e_3 e_6
\cr &&+1212 e_4 e_6+5016 e_5 e_6+4 e_2 (1938 e_3+1439 e_4+6078 e_5+1314 e_6))+\ell_0^8)
\cr && +\mathcal{O}(e_i^3)\,.\label{s-order2-D=5}
\eea
The effective Newton constant now is
\be
\kappa_{\rm eff}=1-\fft{2(60e_1+8e_2+e_4+4e_5)}{\ell_0^4}-\fft{8(60e_1+8e_2+e_4+4e_5)^2}{3\ell_0^8}+\mathcal{O}(e_i^3)\,,
\ee
yielding the mass density
\bea
m &=& \frac{r_0^4}{\ell_0^{10}} (-4 (1650 e_1+430 e_2+126 e_3+101 e_4+378 e_5+86 e_6) \ell_0^4
\cr && +4 (20239200 e_1^2+30 (380804 e_2+120540 e_3+88195 e_4+380024 e_5+81876 e_6) e_1
\cr &&+1584808 e_2^2+151704 e_3^2+84721 e_4^2+1475568 e_5^2+69544 e_6^2+228042 e_3 e_4
\cr &&+946968 e_3 e_5+714762 e_4 e_5+205440 e_3 e_6+154634 e_4 e_6+640808 e_5 e_6
\cr &&+e_2 (987840 e_3+732850 e_4+3098376 e_5+669904 e_6))+3 \ell_0^8)+\mathcal{O}(e_i^3)\,.
\eea
It is easy to see (\ref{1slaw}) is valid. Moreover, we can check immediately
\be
m=\fft{3}{4}T s\,,
\ee
which should hold for a thermal plasma in $d=4$ CFT. Notice the vacuum is (\ref{vac-5D-order-1}) with (\ref{2-ord-ell-5D}), the Free energy density is given by
\bea
\mathcal{F}&=&-\frac{r_0^4}{3 \ell_0^{10}} (-4 (1650 e_1+430 e_2+126 e_3+101 e_4+378 e_5+86 e_6) \ell_0^4
\cr && +4 (20239200 e_1^2+30 (380804 e_2+120540 e_3+88195 e_4+380024 e_5+81876 e_6) e_1
\cr &&+1584808 e_2^2+151704 e_3^2+84721 e_4^2+1475568 e_5^2+69544 e_6^2+228042 e_3 e_4
\cr &&+946968 e_3 e_5+714762 e_4 e_5+205440 e_3 e_6+154634 e_4 e_6+640808 e_5 e_6
\cr && +e_2 (987840 e_3+732850 e_4+3098376 e_5+669904 e_6))+3 \ell_0^8)+\mathcal{O}(e_i^3)\,.
\eea
One can verify (\ref{ther-rela}).

In $D=4$, with the solutions (\ref{f-k=0-D=4-2order}) and (\ref{h-k=0-D=4-2order}), we have the temperature as follows
\bea
T &=& \frac{3 r_0}{32 \pi  \ell_0^{10}} (-12 (12 e_1+6 e_2+3 e_3+2 e_4+4 e_5+2 e_6) \ell_0^4+3 (398736 e_1^2
\cr &&+12 (19776 e_2+6525 e_3+5039 e_4+16640 e_5+4350 e_6) e_1+34524 e_2^2
\cr &&+3402 e_3^2+2242 e_4^2+22256 e_5^2+1512 e_6^2+5631 e_3 e_4+17340 e_3 e_5
\cr &&+14316 e_4 e_5+4536 e_3 e_6+3754 e_4 e_6+11560 e_5 e_6+6 e_2 (3687 e_3+2931 e_4
\cr && +9400 e_5+2458 e_6))+8 \ell_0^8)+\mathcal{O}(e_i^3)\,.\label{T-order2-D=4}
\eea
The entropy density is given by
\bea
s &=& -2 \pi  r_0^2 (-\frac{54}{\ell_0^8} (11520 e_1^2+48 (144 e_2+48 e_3+37 e_4+120 e_5+32 e_6) e_1+1008 e_2^2
\cr && +99 e_3^2+66 e_4^2+640 e_5^2+44 e_6^2+165 e_3 e_4+504 e_3 e_5+416 e_4 e_5+132 e_3 e_6
\cr && +110 e_4 e_6+336 e_5 e_6+12 e_2 (54 e_3+43 e_4+136 e_5+36 e_6))
\cr && +\frac{9}{\ell_0^4} (48 e_1+12 e_2+3 e_3+3 e_4+8 e_5+2 e_6)-2)+\mathcal{O}(e_i^3)\,.\label{s-order2-D=4}
\eea
Notice we have
\be
\kappa_{\rm eff}=1-\fft{2(36e_1+6e_2+e_4+4e_5)}{\ell_0^4}-\fft{8(36e_1+6e_2+e_4+4e_5)^2}{3\ell_0^8}+\mathcal{O}(e_i^3)\,,
\ee
we immediately have the mass density
\bea
m &=& \frac{r_0^3}{4 \ell_0^{10}} (-12 (156 e_1+42 e_2+12 e_3+11 e_4+28 e_5+8 e_6) \ell_0^4+3 (1238544 e_1^2
\cr && +12 (61896 e_2+20619 e_3+15893 e_4+51632 e_5+13746 e_6) e_1+108396 e_2^2
\cr && +10692 e_3^2+7102 e_4^2+68912 e_5^2+4752 e_6^2+17781 e_3 e_4+54276 e_3 e_5
\cr &&+44772 e_4 e_5+14256 e_3 e_6+11854 e_4 e_6+36184 e_5 e_6
\cr &&+6 e_2 (11625 e_3+9249 e_4+29272 e_5+7750 e_6))+8 \ell_0^8)+\mathcal{O}(e_i^3)\,.
\eea
It is not difficult to verify that (\ref{1slaw}) is valid. We can also verify
\be
m=\fft{2}{3}Ts\,,
\ee
which is supposed to be valid for a plasma in $d=3$ CFT. Note we have the AdS vacuum given by (\ref{vac-5D-order-1}) with (\ref{2-ord-ell-4D}), hence the Free energy density is
\bea
\mathcal{F}&=&-\frac{r_0^3}{8 \ell_0^{10}} (-12 (156 e_1+42 e_2+12 e_3+11 e_4+28 e_5+8 e_6) \ell_0^4+3 (1238544 e_1^2
\cr && +12 (61896 e_2+20619 e_3+15893 e_4+51632 e_5+13746 e_6) e_1+108396 e_2^2
\cr && +10692 e_3^2+7102 e_4^2+68912 e_5^2+4752 e_6^2+17781 e_3 e_4+54276 e_3 e_5
\cr &&+44772 e_4 e_5+14256 e_3 e_6+11854 e_4 e_6+36184 e_5 e_6+6 e_2 (11625 e_3
\cr && +9249 e_4+29272 e_5+7750 e_6))+8 \ell_0^8)+\mathcal{O}(e_i^3)\,.
\eea
We still have (\ref{ther-rela}).
\section{Central charges and two-point functions}
\label{2-pt}
The holographic central charges and the holographic two-point function of the energy-momentum tensor for the cubic gravities were studied extensively in the literature \cite{Myers:2010jv,Peng:2018vbe,deBoer:2009gx,Li:2017txk,Li:2018drw}. In this section, we review those results for the most generic massless cubic gravities \cite{Li:2017txk,Li:2018drw}. We start with the holographic conformal anomaly in $D=5$, $d=4$. The conformal anomaly exhibits its universal structure in $d=4$ as
\bea
\mathcal{A}_{\rm anom} = -a E^{(4)} + c I^{(4)}\,,\label{D5anom}
\eea
where $E^{(4)}$ is the Euler density in $d=4$ and $I^{(4)}$ is the Weyl invariants in $d=4$, they are given by
\be
E^{(4)} = \mathcal{R}^2 - 4 \mathcal{R}^{ij} \mathcal{R}_{ij} + \mathcal{R}^{ijkl} \mathcal{R}_{ijkl}\,,\qquad
I^{(4)} = \ft13 \mathcal{R}^2-2\mathcal{R}_{ij}\mathcal{R}^{ij}+\mathcal{R}_{ijkl}\mathcal{R}^{ijkl}\,.
\ee
There arises two coefficients in the conformal anomaly (\ref{D5anom}) and they are referred as $a$-central charge and $c$-central charge respectively. It turns out that the universal information of a CFT is encoded in the central charges. Specifically, $a$-charge measures the massless freedom, and one can find a corresponding $a$-function that gives rise to the $a$-charge at fixed point and encodes the RG flow properties. On the other hand, $c$-charge is related to the universal coefficient of the energy-momentum tensor two-point function in $d=4$ which shall be discussed momentarily.

Holographically, it turns out that even though the conformal anomaly only appears in odd bulk dimension $D$, one can always generalize the $a$-charge to arbitrary dimensions \cite{Li:2017txk,Myers:2010xs,Myers:2010tj}. In fact, with the reduced FG expansion trick in \cite{Li:2017txk,Li:2018kqp} one can readily read off the $a$-charge in arbitrary odd $D$, even $d$. Then one can always impose the analytic continuation for $a$-charge and state there are certain $a$-charges in arbitrary dimensions. From another perspective, the $a$-function\footnote{sometimes, it is also referred as $c$-function or $C$-function in the literatures} and the corresponding holographic $a$-theorem is related to the null energy condition \cite{Freedman:1999gp,Alkac:2018whk,Li:2017txk,Myers:2010xs,Myers:2010tj,Li:2018kqp} that is intact in any dimensions, hence at the fixed point, the $a$-function automatically gives rise to the $a$-charge in arbitrary dimensions. The $a$-charge of the massless cubic gravities in general dimensions are given by \cite{Li:2017txk}
\be
a=\fft{2\pi^{\fft{d}{2}}}{\Gamma(\fft{d}{2})}\ell^{d-1}\Big(1+\fft{1}{\ell^4}(d-2)(d-1)(3(d+1)de_1+2de_2+e_4+4e_5)\Big)\,.\label{a-charge-d}
\ee
The $a$-charge specializes in $D=5$, $d=4$ is given by
\be
a=2\pi^2(\ell^3 +6(60 e_1 + 8 e_2 + e_4 + 4 e_5)\ell^{-1})\,.\label{a-5D}
\ee
In even $D$, i.e. odd $d$, it turns out the so called $a$-charge is related to the coefficient of the entanglement entropy \cite{Myers:2010xs,Myers:2010tj,Hung:2011xb,Myers:2012ed}. We will see in section \ref{entropy} that this fact holds true in the massless cubic gravities. We also present the $c$-charge in $D=5$, $d=4$ \cite{Li:2017txk}
\be
c=2\pi^2(\ell^3 - 2(60 e_1 + 8 e_2 + e_4 + 4 e_5)\ell^{-1})\,,\label{c-5D}
\ee
By writing the excited modes around the AdS vacua (i.e. the modes that are solutions of (\ref{Newtoneff})) in the metric basis, see e.g. \cite{Li:2018drw,Johansson:2012fs,Ghodsi:2014hua} and also \cite{Li:2018rgn}), one can prove that the holographic energy-momentum tensor two-point function of the massless cubic gravities in general dimensions takes the form as \cite{Li:2018drw}
\be
\langle T_{ij}(x)T_{kl}(0)\rangle=\fft{C_T\,
\mathcal{I}_{ijkl}(x)}{x^{2d}}\,,
\ee
where $\mathcal{I}_{ijkl}(x)$ is defined as
\bea
\mathcal{I}_{ijkl}(x)=\ft{1}{2}\big(I_{ik}(x)I_{jl}(x)+I_{il}(x)I_{jk}(x)\big)-
\ft{1}{d}\eta_{ij}\eta_{kl}\,,\qquad I_{ij}(x)=\eta_{ij}-\fft{2x_{i}x_{j}}{x^2}\,,\label{Iijkl}
\eea
and $C_T$ is given by
\bea
C_T=\ft{2\Gamma(d+2)}{\pi^{\fft{d}2} (d-1)\Gamma(\fft{d}2)}\ell^{d-1}\Big(1+\fft{1}{\ell^4}(d-5)(d-2)(3(d+1)de_1+2de_2+e_4+4e_5)\Big)\,.
\eea
In $D=5$, $d=4$, one can immediately observe that it is proportional to the $c$-charge \cite{Li:2018drw}
\be
C_T=\fft{40}{\pi^4}c\,.\label{CT-c-rela}
\ee
Sometimes it would be convenient to strip off the inessential numerical factors $N_1$, $N_2$ in $a=N_1 \tilde{a}$, $c=N_1 \tilde{c}$ and $C_T=N_2 \mathcal{C}_T$ such that we have \cite{Li:2017txk,Li:2018drw}
\bea
&& \tilde{a}=\ell^{d-1}\Big(1+\fft{1}{\ell^4}(d-2)(d-1)(3(d+1)de_1+2de_2+e_4+4e_5)\Big)\,,
\cr &&
\cr && \tilde{c}|_{d=4}=\mathcal{C}_T|_{d=4}=(\ell^3 - 2(60 e_1 + 8 e_2 + e_4 + 4 e_5)\ell^{-1})\,,
\cr &&
\cr && \mathcal{C}_T=\ell^{d-1}\Big(1+\fft{1}{\ell^4}(d-5)(d-2)(3(d+1)de_1+2de_2+e_4+4e_5)\Big)\,.\label{a-c-CT}
\eea
Actually, it would be convenient to identify $a$ with $\tilde{a}$ and $c$ with $\tilde{c}$ when there is no confusion since the numerical factor is inessential for the theory detail \cite{Li:2017txk,Li:2018drw} (indeed, in determining parameters of three-point functions, the numerical factors should be no longer ignored, see \cite{Buchel:2009sk,Myers:2010jv} and section \ref{flux}). Moreover, there is a remarkable relation between $a$-charge and $\mathcal{C}_T$ that was found recently in \cite{Li:2018drw}
\be
\mathcal{C}_T=\fft{1}{d-1}\,\ell \fft{\partial \tilde{a}}{\partial \ell}\,.\label{CT-uni}
\ee
An equivalent relation between $C_T$ and the Free energy of CFT on a sphere was also observed recently in \cite{Bueno:2018yzo}. In particular, in $d=4$ one has
\be
\tilde{c}=\fft{1}{3}\ell \fft{\partial \tilde{a}}{\partial \ell}\,,\label{c-a}
\ee
which amazingly connects two types of central charges via the derivative relation over the effective AdS radius. It is of great interests to investigate the fascinating relation (\ref{CT-uni}) and (\ref{c-a}) directly in CFT. Without the numerical factor, we have $\tilde{c}=\mathcal{C}_T$ in $D=5$, $d=4$, for this reason, we may also refer $\mathcal{C}_T$ as the $c$-charge in other dimensions and then $\mathcal{C}_T$ takes the responsibility as $c$ in this paper.
\section{The energy flux}
\label{flux}
\subsection{The energy flux parameters}
In this section, we mainly focus on completing the holographic dictionary of the massless cubic gravities in $D=5$. One should note that in this section, we do not slip off the numerical factor in front of central charges and $C_T$.

In the previous section, we state that $a$-charge and $c$-charge ($C_T$-charge to be precise) are universal for a certain CFT to determine itself. Specifically, the $a$-charge encodes the RG feature, meanwhile $c$-charge determines the two-point function of energy-momentum tensor. Even though in the holographic context, we only discuss the pure gravity which only provides the information associated with the energy-momentum tensor in the boundary CFT, it is still far from enough to determine all the information regarding the dynamics that energy-momentum tensor is solely responsible for. There are three independent universal parameters exist in the three-point functions of the energy-momentum tensor, see, e.g. \cite{Osborn:1993cr,Erdmenger:1996yc}. It turns out that the conformal invariance can be used to explicitly determine the five independent structure of the energy-momentum tensor three-point function, hence one should have five universal parameters carved with the CFT information \cite{Osborn:1993cr,Erdmenger:1996yc}. Considering the conserved law of the energy-momentum tensor, the total independent parameters are three, which are denoted as $\mathcal{A}$, $\mathcal{B}$ and $\mathcal{C}$ in \cite{Erdmenger:1996yc}. One should expect that we can manage to obtain the three-point function and hence the parameters holographically, however, unfortunately, it is a tremendously difficult task even for Einstein gravity (see \cite{Arutyunov:1999nw} for the calculation of Einstein gravity). Therefore, we shall seek other ways to determine the parameters $\mathcal{A}$, $\mathcal{B}$ and $\mathcal{C}$.

We follow the discussion for quasi-topological gravity and Lovelock gravities in \cite{Myers:2010jv,Buchel:2009sk} to consider the conformal collider thought experiment proposed in \cite{Hofman:2008ar}. Specifically, consider we put a local excitation $\mathcal{O}$ in a $d$ dimensional CFT, the local excitation would spread out in the spacetime, then one would like to measure the resulting energy flux along the null infinity flowing in the direction $n^i$. It is instructive to consider the energy flux operator in the direction $n^i$
\be
\mathcal{E}(n^i)=\lim_{r\rightarrow\infty}r^{d-2}\int_{-\infty}^{+\infty}dt\, T^t\,_i(t,r\,n^i)n^i\,.
\ee
Then the one-point function of the energy flux operator can be obtained by evaluating the three-point function with respect to the vacuum as follows
\be
\langle\mathcal{E}(n^i)\rangle=\fft{\langle\mathcal{O}^\dagger\mathcal{E}(n^i)\mathcal{O}\rangle}
{\langle\mathcal{O}^\dagger\mathcal{O}\rangle}\,.\label{energy-flux-defin}
\ee
It is convenient to work in the light-cone cooridates, i.e.
\be
ds^2=-dx^{+}dx^{-}+dx_{\tilde{i}}dx^{\tilde{i}}\,,\label{light-cone1}
\ee
where $\tilde{i}$ only covers the last $d-2$ directions in $i$, i.e. $\tilde{i}=(2,3,\cdots,d-1)$, $x^{+}$ and $x^{-}$ are the holomorphic and anti-holomorphic coordinates given by
\be
x^{+}=t+x_1\,,\qquad x^{-}=t-x_1\,.
\ee
Furthermore, the following coordinates transformation
\be
y^{+}=-\fft{1}{x^{+}}\,,\qquad y^{-}=x^{-}-\fft{x_{\tilde{i}} x^{\tilde{i}}}{x^{+}}\,,\qquad y^i=\fft{x^{\tilde{i}}}{x^{+}}\,,
\ee
accompanied with the conformal transformation $g_{\mu\nu}\rightarrow (y^{+})^2 g_{\mu\nu}$ leading us to the following coordinates
\be
ds^2=-dy^{+}dy^{-}+dy^{\tilde{i}}dy_{\tilde{i}}\,,\label{light-cone2}
\ee
in which we end up with a simpler formula to compute the energy flux operator
\be
y^{\tilde{i}}=\fft{n^{\tilde{i}}}{1+n^{d-1}}\,,\qquad \mathcal{E}(n^i)=\Omega^{d-1}\int_{-\infty}^{+\infty}dy^{-}T_{--}(y^{+}=0,
y^{-},y^{\tilde{i}})\,,\label{energy-flux}
\ee
where $\Omega$ is given by
\be
\Omega=\fft{1}{1+n^{d-1}}\,.\label{factor}
\ee
In this coordinate, the energy flux is measured in the surface of $y^{+}=0$ which is the future null infinity. To relate the energy flux with the three-point functions of energy-momentum tensor, we shall consider the operator $T_{ij}\varepsilon^{ij}$ as the excitation, where $\varepsilon_{ij}$ is the constant polarization tensor. The symmetries of the construction can determine two independent structures with two independent parameters $t_2$ and $t_4$
\be
\langle\mathcal{E}\rangle=\fft{E}{\omega_{d-2}}\Big(1+t_2\big(\fft{\varepsilon^{\ast}_{ij}\varepsilon^{i}_k n^j n^k}{\varepsilon^{\ast}_{ij}\varepsilon^{ ij}}-\fft{1}{d-1}\big)+t_4\big(\fft{\big|\varepsilon_{ij}n^i n^j\big|^2}{\varepsilon^{\ast}_{ij}\varepsilon^{ij}}-\fft{2}{d^2-1}\big)\Big)\,,\label{flux-cft}
\ee
where $E$ is the total energy and $\omega_{d-2}$ is the volume of a unit $(d-2)$-sphere. The energy flux considered here is actually the energy-momentum tensor three-point functions restricted to certain polarization, therefore, it is natural to relate $t_2$ and $t_4$ to $\mathcal{A}$, $\mathcal{B}$ and $\mathcal{C}$. Indeed, it turns out the energy flux parameters $t_2$ and $t_4$ can be expressed in terms of $\mathcal{A}$, $\mathcal{B}$ and $\mathcal{C}$ \cite{Myers:2010jv,Buchel:2009sk,Hofman:2008ar}. In fact, with using the Ward identity, one can express $C_T$ in terms of $\mathcal{A}$, $\mathcal{B}$ and $\mathcal{C}$ by \cite{Osborn:1993cr,Erdmenger:1996yc}
\be
C_T=\fft{\pi^{\fft{d}{2}}}{\Gamma(\fft{d}{2})}\fft{(d-1)(d+2)\mathcal{A}-2\mathcal{B}-4(d+1)\mathcal{C}}{d(d+2)}\,.\label{Ward}
\ee
Therefore, with the two-point functions coefficient $C_T$ and the energy flux parameters $t_2$, $t_4$, one could, in principle, solve out $\mathcal{A}$, $\mathcal{B}$ and $\mathcal{C}$. In this sense, $t_2$ and $t_4$ are the universal parameters we found.

Before we explicitly calculate the holographic energy flux parameters $t_2$ and $t_4$ for $D=5$ massless cubic gravities, it is worth noting that in $d=3$, we only have $t_4$ parameter
\be
\langle\mathcal{E}\rangle=\fft{E}{2\pi}\Big(1+t_4\big(\fft{\big|\varepsilon_{ij}n^i n^j\big|^2}{\varepsilon^{\ast}_{ij}\varepsilon^{ij}}-\fft{1}{4}\big)\Big)\,.\label{flux-cft-d=3}
\ee
It would be challenging to obtain $t_4$ for $D=4$ massless cubic gravities by following the procedure in \cite{Myers:2010jv,Buchel:2009sk} and in this section. However, in section \ref{entropy}, we would use the scaling dimension of the twist operator to extract $t_4$ in $D=4$ just like one did in \cite{Bueno:2018xqc}.
\subsection{The holographic energy flux}
We consider the AdS background (\ref{AdS-1}), and to proceed, it is way more convenient to have the boundary metric exactly like (\ref{light-cone2}). We follow \cite{Myers:2010jv,Buchel:2009sk,Hofman:2008ar} to impose the coordinates transformation for AdS vacuum as follows
\bea
y^{+}=-\fft{1}{x^{+}}\,,\qquad y^{-}=x^{-}-\fft{x_{\tilde{i}} x^{\tilde{i}}}{x^{+}}-\fft{\ell^2}{r^2 x^{+}}\,,\qquad y^{\tilde{i}}=\fft{x^{\tilde{i}}}{x^{+}}\,,\qquad \rho=r x^{+}\,.\label{coordinate-transf}
\eea
It can be verified easily that
\bea
ds^2 &=& \fft{\ell^2}{r^2}dr^2+r^2 \eta_{ij}dx^i dx^j
\cr &&
\cr &=& \fft{\ell^2}{r^2}dr^2+r^2 (-dx^{+}dx^{-}+dx^{\tilde{i}}dx_{\tilde{i}})
\cr &&
\cr &=& \fft{\ell^2}{\rho^2}d\rho^2+\rho^2 (-dy^{+}dy^{-}+dy^{\tilde{i}}dy_{\tilde{i}})\,.\label{AdS-coor}
\eea
Even though the $(\rho,y)$ coordinates look exactly the same as the usual AdS coordinates $(r,x)$, it should be noticed that the corresponding energy flux operator constructed in the boundary CFT associated with $(\rho,y)$ coordinates should follow (\ref{energy-flux}). Apparently, from (\ref{energy-flux}), the energy flux operator is the energy-momentum tensor operator along $y^{-}y^{-}$ direction integrated over $y^{-}$, implying that it is in fact an operator with scaling dimension $\Delta=d-1$. According to the holographic dictionary, $T_{--}$ in CFT should be coupled to the perturbation in the AdS background along $g_{++}$, since the energy flux $\mathcal{E}$ in (\ref{energy-flux}) has no dependence on $y^{-}$ and it is localized at $y^{+}$, we should have the perturbation as
\be
ds^2=\fft{\ell^2}{\rho^2}d\rho^2+\rho^2 (-dy^{+}dy^{-}+(dy^1)^2+(dy^2)^2+\delta(y^{+})W(\rho,y^1,y^2)(dy^{+})^2)\,.
\ee
As we mentioned before, the integration over $y^{-}$ implies that $W(\rho,y^1,y^2)$ couples to an operator with scaling dimension $\Delta=d-1$, in this case, $\Delta=3$, therefore, one can immediately write down the solution for the propogator $W$
\be
W(\rho,y^1,y^2)\sim \fft{1}{\rho^4\Big((y^1-y^{'1})^2+(y^2-y^{'2})^2+\ell^2/\rho^2\Big)^3}\,,\label{W-solu}
\ee
where the overall normalized factor is inessential for our purpose and so we slip it off for simplicity, and $y^{'1}$ and $y^{'2}$ represents the direction where we detect the energy flux, i.e. from the first formula in (\ref{energy-flux})
\be
y^{'1}=\fft{n^1}{1+n^3}\,,\qquad y^{'2}=\fft{n^2}{1+n^3}\,,\qquad (n^1)^2+(n^{2})^2+(n^{3})^2=1\,.\label{y-n}
\ee
Indeed, we can verify that \label{W-solu} is the solution of the linearized equation of motion associated with $W$
\be
(\partial_1^2 W+\partial_2^2 W)\ell^2+\rho^3 (5\partial_\rho W+\rho \partial_\rho^2 W)=0\,.\label{W-equi}
\ee
In addition to the energy flux, we should include the local excitation operator $\mathcal{O}=T_{ij}\varepsilon^{ij}$. For convenience, we choose $\varepsilon^{x^1x^2}=\varepsilon^{x^2x^1}=1$ with other components vanishing. Correspondingly we can perturb the metric by $h_{12}(\rho,y^{+},y^{-},y^1,y^2)dy^1 dy^2$. However, even though it seems that only $h_{12}$ contributes to the final answer, it is worth noting that the nontrivial perturbation should ensure the traceless transverse gauge, i.e.
\be
\nabla_\mu h^{\mu\nu}=0\,,\qquad h=0\,,\label{tras-trace}
\ee
which is enforcing us to consider other perturbations that are nontrivial and indeed contribute to the final result (actually, for Gauss-Bonnet gravity they do not contribute. But for the generic gravity, they are indeed nontrivial). The total perturbations forming the full spectrum for our purpose are given by
\bea
ds^2 &=& \fft{\ell^2}{\rho^2}d\rho^2+\rho^2 (-dy^{+}dy^{-}+(dy^1)^2+(dy^2)^2+\delta(y^{+})W(\rho,y^1,y^2)(dy^{+})^2
\cr && +h_{++}(\rho,y^{+},y^{-},y^1,y^2)(dy^{+})^2+2h_{+1}(\rho,y^{+},y^{-},y^1,y^2)dy^{+}dy^1
\cr && +2h_{+2}(\rho,y^{+},y^{-},y^1,y^2)dy^{+}dy^2+2h_{12}(\rho,y^{+},y^{-},y^1,y^2)dy^{1}dy^2)\,.\label{per}
\eea
Other perturbations are decoupled from the spectrum associated with $T_{12}$, and hence we do not turn on them for simplicity. The gauge condition (\ref{tras-trace}) implies that the perturbations without $W$ in (\ref{per}) actually belong to one singlet, namely $h_{12}$, explicitly, we have
\be
\partial_{-}h_{+1}=\fft{1}{2}\partial_2 h_{12}\,,\qquad \partial_{-}h_{+2}=\fft{1}{2}\partial_1 h_{12}\,,\qquad \partial_{-}^2h_{++}=
\fft{1}{2}\partial_1 \partial_2 h_{12}\,.\label{h-relation}
\ee
Then the equation of motion gives rise to one nontrivial equation
\be
(\partial_1^2 h_{12}+\partial_2^2 h_{12}+4\partial_{+} \partial_{-} h_{12})\ell^2+5\rho^3 \partial_\rho h_{12}+\rho^4 \partial_\rho^2 h_{12}=0\,.\label{h12-eq}
\ee
The equation (\ref{h12-eq}) just gives rise to the propogator corresponds to the $\Delta=4$ energy-momentum tensor operator, and in coordinates $(r,x)$, the propogator is given by (\ref{prop}) in Appendix \ref{local} provided $\Delta=4$. For our purpose, the necessary information is not the propogator solution but the localized property. It turns out that the corresponding propogator dual to the excitation operator in $(\rho,y)$ coordinates is localized at $\rho=\ell$ and $y^1=y^2=0$, see, e.g \cite{Myers:2010jv,Hofman:2008ar} or Appendix \ref{local} for a brief sketch. Then we substitute our on-shell perturbations (\ref{per}) into the bulk action $S_{\rm bulk}$, recall that our purpose is to read off the coefficient appears in the three-point functions of $\langle T_{ij}\varepsilon^{ij}\mathcal{E}T_{ij}\varepsilon^{ij}\rangle$, we shall extract the terms as the effective action such that it has quadratic $h$-type perturbations and linear $W$ perturbation. We would like to employ the transverse-traceless gauge condition (\ref{h-relation}), subsequently we find there would only involve the on-shell perturbations $h_{12}$ and $W$ in the effective action while other $h$-type perturbations do not appear anymore. By integrating by parts and applying the linearized equations of motion (\ref{W-equi}) and (\ref{h12-eq}) (recalling $h_{12}$ is localized at $\rho=\ell$ and $y^1=y^2=0$), the effective action takes the form as
\bea
S_{\rm eff} &=& -\fft{4}{\ell^2}\int d^5x\,\delta(y^{+})h_{12}\, \partial_{-}^2 h_{12}W \Big(\mathcal{C}_T\,\rho+\fft{\alpha_1}{\rho}\fft{(\partial_1^2 W+\partial_2^2 W)\ell^2+2\rho^3 \partial_\rho W}{W}
\cr &&
\cr && +\fft{\alpha_2}{\rho^3}\fft{(\partial_\rho\partial_1^2W+\partial_\rho\partial_2^2W)\rho^3+\partial_1^2 \partial_2^2 W \ell^2}{W}\Big)\Big|_{\rho=\ell,y^1=y^2=0}\,,\label{S3-cubic}
\eea
where $\alpha_1$ and $\alpha_2$ are given by
\bea
&& \alpha_1=\fft{4(270e_1+66e_2+18e_3+15e_4+38e_5+10e_6)}{\ell}\,,
\cr && \alpha_2=-2(600e_1+140e_2+36e_3+31e_4+80e_5+20e_6)\ell\,,
\eea
and $\mathcal{C}_T$ is given in the second line of (\ref{a-c-CT}) in section \ref{2-pt}. We now can read off the coefficient $C_{hhW}$ of the three-point function $\langle T_{ij}\varepsilon^{ij}\mathcal{E}T_{ij}\varepsilon^{ij}\rangle$, slipping off the inessential numerical factors and using (\ref{W-solu}) and (\ref{y-n}), we have
\be
\fft{C_{hhW}}{\mathcal{C}_T}=1+t_2(\fft{n_1^2+n_2^2}{2}-\fft{1}{3})+t_4 (2n_1^2 n_2^2-\fft{2}{15})\,,\label{ChhW}
\ee
where
\bea
&& t_2=\fft{48(2340e_1+552e_2+144e_3+123e_4+316e_5+80e_6)}{\ell^4-2(60e_1+8e_2+e_4+4e_5)}\,,
\cr && t_4=-\fft{360(600e_1+140e_2+36e_3+31e_4+80e_5+20e_6)}{\ell^4-2(60e_1+8e_2+e_4+4e_5)}\,.\label{t2-t4}
\eea
Note we have
\be
\fft{\varepsilon^{\ast}_{ij}\varepsilon^{i}_k n^j n^k}{\varepsilon^{\ast}_{ij}\varepsilon^{ ij}}=\fft{n_1^2+n_2^2}{2}\,,\qquad \fft{\big|\varepsilon_{ij}n^i n^j\big|^2}{\varepsilon^{\ast}_{ij}\varepsilon^{ij}}=2n_1^2 n_2^2\,,
\ee
then from (\ref{flux-cft}) specialized in $d=4$ we have
\be
\langle\mathcal{E}\rangle=\fft{E}{\omega_{2}}\Big(1+t_2\big(\fft{n_1^2+n_2^2}{2}-\fft{1}{3}\big)+t_4\big(2n_1^2 n_2^2-\fft{2}{15}\big)\Big)\,.\label{flux-cft2}
\ee
Compare (\ref{ChhW}) with (\ref{flux-cft2}), we can conclude that $t_2$ and $t_4$ in (\ref{t2-t4}) are indeed the energy flux parameters we want to obtain.

For Myers quasi-topological gravity specialized in $D=5$ \cite{Myers:2010ru,Myers:2010jv} (we follow the notation in \cite{Myers:2010ru,Myers:2010jv})
\bea
(e_1,e_2,e_3,e_4,e_5,e_6,e_7,e_8)=\fft{7\mu\ell_0^4}{8}\,\Big(\fft{11}{14},-\fft{54}{7},\fft{64}{7},\fft{72}{7},\fft{3}{2},-\fft{60}{7},1,0\Big)\,,
\label{Myers-quasi}
\eea
Following \cite{Myers:2010ru,Myers:2010jv}, we introduce $f_{\infty}=\ell_0^2/\ell^2$, then our results (\ref{t2-t4}) give rise to
\be
t_2=-\fft{2088\mu f_{\infty}^2}{1-3\mu f_{\infty}^2}\,,\qquad t_4=\fft{3780\mu f_{\infty}^2}{1-3\mu f_{\infty}^2}\,,
\ee
which coincides with the results obtained in \cite{Myers:2010jv}.
\subsection{Three-point function parameters and physical constraints}
In this subsection, we shall obtain the three-point function parameters $\mathcal{A}$, $\mathcal{B}$ and $\mathcal{C}$. At first, we should note that the Ward identity can provide us the relation between $C_T$ and $\mathcal{A}$, $\mathcal{B}$ and $\mathcal{C}$, i.e. (\ref{Ward}). Specializing in $d=4$ yields
\be
C_T=\fft{\pi^2}{12}(9\mathcal{A}-\mathcal{B}-10\mathcal{C})\,,\qquad C_T=\fft{80}{\pi^2}(\ell^3 - 2(60 e_1 + 8 e_2 + e_4 + 4 e_5)\ell^{-1})\,.\label{CT-ABC}
\ee
In addition, we have \cite{Myers:2010jv,Buchel:2009sk,Camanho:2013pda}
\be
t_2=\frac{15 (5 \mathcal{A}+4 \mathcal{B}-12 \mathcal{C})}{9 \mathcal{A}-\mathcal{B}-10 \mathcal{C}}\,,\qquad t_4=-\frac{15 (17 \mathcal{A}+32 \mathcal{B}-80 \mathcal{C})}{4 (9 \mathcal{A}-\mathcal{B}-10\mathcal{C})}\,,\label{t2-t4-ABC}
\ee
then from (\ref{CT-ABC}) and (\ref{t2-t4-ABC}), we can solve
\bea
&& \mathcal{A}=\frac{1024 (4 (30 e_1+34 e_2+18 e_3+11 e_4+22 e_5+10 e_6)-\ell^4)}{9 \pi ^4 \ell}\,,
\cr && \mathcal{B}=\frac{64 (2 (100140 e_1+24392 e_2+6624 e_3+5533 e_4+14036 e_5+3680 e_6)-49 \ell^4)}{9 \pi ^4 \ell}\,,
\cr && \mathcal{C}=-\frac{64 (2 (8340 e_1+1352 e_2+144 e_3+223 e_4+716 e_5+80 e_6)+23 \ell^4)}{9 \pi ^4 \ell}\,.\label{ABC}
\eea
One can also immediately verify some other identities \cite{Myers:2010jv,Buchel:2009sk,Hofman:2008ar}
\be
a=\frac{\pi ^6 (13 \mathcal{A}-2 \mathcal{B}-40 \mathcal{C})}{2880}\,,\qquad 1-\fft{a}{c}=\fft{1}{6}t_2+\fft{4}{45}t_4\,.
\ee
For Myers quasi-topological gravity (\ref{Myers-quasi}), it is straightforward to check that we can reproduce the results obtained in \cite{Myers:2010jv}.

Even though we have already removed the massive modes, implying the resulting massless gravity theories shall not suffer from ghosts and the dual CFT should be unitary in Lorentzian signature (or reflection positive in Euclidean signature), however, it is easy to see from the negative signs in (\ref{a-c-CT}) (or (\ref{CT-ABC})) and (\ref{flux-cft2}) that coupling constants might violate the unitarity or reflection positivity. Therefore, we must impose some inequality for coupling constants $e_i$ for preserving the unitarity. The first obvious constraint is $C_T\sim c>0$ which requires
\be
60e_1+8e_2+e_4+4e_5<\fft{\ell^4}{2}\,.\label{constraint1}
\ee
Note the requirement (\ref{constraint1}) does not impose any further constraint for $e_6$. In most parts of this paper, e.g. section \ref{cubic-gravity}, section \ref{entropy} and section \ref{hydro}, we treat the coupling constants $e_i$ as infinitesimal quantities compared to any other gravity theory parameters including $\ell_0$ and expand all relevant physical quantities with respect to $e_i$ up to $\mathcal{O}(e_i^2)$ or $\mathcal{O}(e_i^3)$, under this consideration, the constraint (\ref{constraint1}) is trivially satisfied.

Furthermore, we shall consider the constraint by demanding the energy flux is non-negative, i.e. $\langle\mathcal{E}\rangle\geq0$. It turns out this requirement actually imposes three constraints, see \cite{Myers:2010jv,Buchel:2009sk,Camanho:2013pda}. We follow \cite{Buchel:2009sk} to classify the constraints as "tensor channel", "vector channel" and "scalar channel". The classification scheme is reviewed as follows: For convenience we set $n^i=(1,0,0)$, we classify the channels according to the transformation properties of $\varepsilon_{ij}$ under the ${\rm SO}(3)$ group that leaves $n^i$ invariant. Explicitly, for $n^i=(1,0,0)$ we have the group elements $g\in{\rm SO}(3)$ taking the form as
\be
g=\left(
\begin{array}{ccc}
 1 & 0 & 0 \\
 0 & \cos\theta & \sin\theta \\
 0 & -\sin\theta & \cos\theta \\
\end{array}
\right)\,.
\ee
Consider $\varepsilon_{23}=\varepsilon_{32}=1$ with other component vanishing, the polarization tensor $\varepsilon_{ij}$ transforms as a tensor, i.e.
\be
g\left(
\begin{array}{ccc}
 0 ~~~& 0 ~~~& 0 \\
 0 ~~~& 0 ~~~& 1 \\
 0 ~~~& 1 ~~~& 0 \\
\end{array}
\right)g^{-1}=
\left(
\begin{array}{ccc}
 0 & 0 & 0 \\
 0 & \sin(2\theta) & \cos(2\theta) \\
 0 & \cos(2\theta) & -\sin(2\theta)\\
\end{array}
\right)\,,
\ee
the corresponding constraint is classified as the tensor channel; consider $\varepsilon_{12}=\varepsilon_{21}=1$ with other component vanishing, the polarization tensor $\varepsilon_{ij}$ transforms as a vector, i.e.
\be
g\left(
\begin{array}{ccc}
 0 ~~~& 1 ~~~& 0 \\
 1 ~~~& 0 ~~~& 0 \\
 0 ~~~& 0 ~~~& 0 \\
\end{array}
\right)g^{-1}=
\left(
\begin{array}{ccc}
 0 & \cos\theta & -\sin\theta \\
 \cos\theta & 0 & 0 \\
 -\sin\theta & 0 & 0 \\
\end{array}
\right)\,.
\ee
the corresponding constraint is classified as the vector channel; in the end, for $\varepsilon_{ij}=1/\sqrt{3}\,{\rm dig}(-2,1,1)$, it is invariant under ${\rm SO}(3)$
\be
g\, \fft{1}{\sqrt{3}}{\rm dig}(-2,1,1)\, g^{-1}=\fft{1}{\sqrt{3}}{\rm dig}(-2,1,1)\,,
\ee
hence it is classified as the scalar channel. In conclusion, we shall have three independent constraints
\begin{itemize}
\item[1.] Tensor channel

\be
1-\fft{1}{3}t_2-\fft{2}{15}t_4\geq0\,.
\ee

\item[2.] Vector channel

\be
1+\fft{1}{6}t_2-\fft{2}{15}t_4\geq0\,.
\ee

\item[3.] Scalar channel

\be
1+\fft{1}{3}t_2+\fft{8}{15}t_4\geq0\,.
\ee
\end{itemize}
Specifically, together with (\ref{constraint1}) we have the constraints as
\bea
&& 4380e_1+1064 e_2+288 e_3+241 e_4+612 e_5+160e_6\leq \fft{\ell^4}{2}\,,
\cr && 4740e_1+1112 e_2+288 e_3+247 e_4+636e_5+160e_6\geq-\fft{1}{10}\ell^4\,,
\cr && 38940e_1+9032e_2+2304e_3+1933e_4+5156e_5+1280e_6\leq\fft{\ell^4}{2}\,.\label{constraint2}
\eea
From the physical constraints (\ref{constraint1}) and (\ref{constraint2}), we can conclude that treating $e_i$ as very small quantities throughout this paper admits a large extent of safety to guarantee the unitarity and the positivity of energy for the dual CFT.

Before ending this section, it is necessary to comment the physical constraints in $D=4$. Similarly, in $D=4$, $d=3$, we have the constraints
\be
\mathcal{C}_T|_{d=3}>0\,,\qquad -4\leq t_4\leq 4\,.\label{D=4-constriant}
\ee
The explicit physical constraints in $D=4$ would be presented in section \ref{entropy} after $t_4$ being obtained from the scaling dimension of the twist operators.
\section{R\'enyi entropy}
\label{entropy}
Entanglement is very fundamental and important in physics, e.g. \cite{Levin:2006zz,Srednicki:1993im,Calabrese:2004eu,Ryu:2006bv}. The entanglement entropy (EE) serves as a very powerful tool to measure the entanglement between different parts of one system, see, e.g. \cite{Levin:2006zz,Srednicki:1993im} and the references therein. More precisely, consider a subsystem $A$ with a density matrix $\rho_A$, then the corresponding EE is defined as the von Neumann entropy associated with the density matrix $\rho_A$, i.e. $S_{\rm EE}=-{\rm Tr}(\rho_A\log\rho_A)$. The concept of EE can be generalized to R\'enyi entropy \cite{Renyi1,Renyi2} which includes more information and is able to reveal more details of the entanglement emerging in the system. A real and positive number $q$ is introduced in R\'enyi entropy such that we have
\be
S_{q}=\fft{1}{1-q}\log{\rm Tr}(\rho_A^q)\,.\label{Renyi}
\ee
In principle, the R\'enyi entropy can provide the full entanglement spectrum \cite{Cala}, and in particular, one can reproduce EE by taking $q\rightarrow 1$, i.e. $S_{\rm EE}=\lim_{q\rightarrow1}S_{q}$. EE and R\'enyi entropy become increasingly important even in the context of AdS/CFT since it is convincing that they might shed a light on the quantum structure of spacetime (see, e.g. \cite{VanRaamsdonk:2009ar,VanRaamsdonk:2010pw}). Therefore, for the completion of the holographic studies of the massless cubic gravities, it is necessary to discuss the holographic R\'enyi entropy with respect to the ground state for the massless cubic gravities.

In this section, we follow the methods developed in \cite{Hung:2011nu,Casini:2011kv} to calculate the holographic R\'enyi entroy for the massless cubic gravities with the approximate black holes up to the first order of $e_i$ in $D=5$ and $D=4$ respectively (i.e. $d=4$ and $d=3$). We also take use of the scaling dimension of the twist operator to verify our results of $t_2$ and $t_4$ (\ref{t2-t4}) for $D=5$ massless cubic gravities, and moreover, to compute $t_4$ parameter for $D=4$ massless cubic gravities.
\subsection{The holographic R\'enyi entropy}
Originally, in order to compute the R\'enyi entropy, one should adopt the "replica trick" to calculate the partition function on a $q$-fold cover of the background geometry \cite{Calabrese:2004eu}, which in fact brings up some difficulties especially in the context of AdS/CFT \cite{Fursaev:2006ih,Headrick:2010zt}. Fortunately, a simpler method of calculating EE was proposed in \cite{Casini:2011kv} and immediately it was generalized to apply to the R\'enyi entropy in \cite{Hung:2011nu}. Essentially, by means of this method, the calculation of EE and R\'enyi entropy spanned with a sphere spacetime region with the radius $\tilde{r}$ for a $d$-dimensional CFT is equivalent to the calculation of thermal entropy supported by the temperature $T=1/(2\pi \tilde{r})$ associated with the hyperbolic cylinder $\mathbb{R}\times \mathbb{H}^{d-1}$ in which the scalar curvature $R$ of the hyperbolic space is given by $R=-(d-1)(d-2)/\tilde{r}^2$. Precisely, note the matrix density of a thermal state is given by
\be
\rho_{\rm ther}=\fft{e^{-\fft{H}{T_0}}}{Z(T_0)}\,,\qquad Z(T_0)={\rm Tr}(e^{-\fft{H}{T_0}})\,.
\ee
The mapping relates the density matrix of the area $A$, i.e. $\rho_A$ to $\rho_{\rm ther}$ by unitary transformation, $\rho_A=U\, \rho_{\rm ther}\, U^{-1}$, then we have
\be
\rho_A^q=U \fft{e^{-q\fft{H}{T_0}}}{Z(T_0)^q}U^{-1}\,,\qquad {\rm Tr}(\rho_A^q)=\fft{Z(\fft{T_0}{q})}{Z(T_0)^q}\,.
\ee
By noting $Z=e^{-\fft{F}{T}}$ (see (\ref{F-partition})) and the thermodynamic relation (\ref{ther-rela}), using (\ref{Renyi}) yields
\be
S_{q}=\fft{q}{1-q}\fft{1}{T_0}\Big(F(T_0)-F(\fft{T_0}{q})\Big)=\fft{q}{1-q}\fft{1}{T_0}\int_{T_0/q}^{T_0}S_{\rm therm}(T)dT\,.\label{Sq}
\ee
The formula (\ref{Sq}) can be used to calculate the holographic R\'enyi entropy with a large extent of simplification. For the purpose of employing (\ref{Sq}) to calculate the holographic R\'enyi entropy, one is supposed to find the black holes solutions with hyperbolic boundary topology. Then, one can readily obtain the black hole entropy by using the Wald formula (\ref{Wald1}), with knowing the explicit black hole entropy as a function of the black hole temperature, evaluating the integral in (\ref{Sq}) immediately can yield the R\'enyi entropy\footnote{It is important to keep in mind that the Wald formula (\ref{Wald1}) is a classical result. While considering the quantum effect, it turns out there would be logarithmic corrections (e.g. \cite{Sen:2012dw}), and consequently the R\'enyi entropy should be modified, see e.g. \cite{Mahapatra:2016iok}.}. This procedure was carried out to obtain the holographic R\'enyi entropy for a large amount of gravity theories including Einstein gravity, Gauss-Bonnet gravity, Myers quasi-topological gravity, Lovelock gravity, Einsteinian cubic gravity and so on \cite{Hung:2011nu,Bueno:2018xqc,Dey:2016pei,Puletti:2017gym,Galante:2013wta,Belin:2013dva}. In this section, we basically follow the procedure to calculate the holographic R\'enyi entropy for the most general massless cubic gravities in $D=5$ and $D=4$ respectively up to the first order of the coupling constants $e_i$ by using the approximate hyperbolic black holes obtained in section \ref{cubic-gravity}. It is worth noting that the holographic R\'enyi entropy calculated in this paper is with respect to pure gravity without considering any matter sectors. Including scalar sectors might cause the instability of the hyperbolic black holes even for Einstein gravity (e.g. \cite{Dias:2010ma,Belin:2013dva}), and this instability is likely to induce a phase transition that can be visualized by the behavior of the R\'enyi entropy \cite{Belin:2013dva,Fang:2016ehk}.

\subsubsection{$D=5$}
At first, it turns out that it would be convenient to introduce the variable $x=r_0\ell$ \cite{Hung:2011nu} where $\ell$ is the effective AdS radius and it is given in (\ref{1-ord-ell-5D}) for $D=5$ approximate black holes of massless cubic gravities. Then we can rewrite (\ref{Sq}) as
\be
S_q=\fft{q}{(q-1)T_0}\Big(S(x)T(x)\big|_{x_q}^1-\int_{x_q}^1 S'(x)T(x)dx\Big)\,,\label{Sq-formu}
\ee
where the prime denotes the derivative with respect to the variable $x$, and $x_q$ is the solution of the equation
\be
T(x_q)=\fft{T_0}{q}\,.\label{xq}
\ee
Note in this prescription, $T_0=T(1)$. For black holes (\ref{f-k-D=5}) and (\ref{h-k-D=5}) with $k=-1$, substituting $r_0=x\ell$ into (\ref{T-5D-1order}) provided $k=-1$ yields
\bea
T(x) &=& -\frac{1}{6 \pi  \ell_0^5 x^5}(x^6 (360 e_1+288 e_2+144 e_3+90 e_4+440 e_5+112 e_6-6 \ell_0^4)
\cr && +x^4 (-300 e_1-400 e_2-216 e_3-131 e_4-644 e_5-168 e_6+3 \ell_0^4)
\cr && +(-720 e_1-96 e_2-12 e_4-48 e_5) x^2+600 e_1+200 e_2+72 e_3+52 e_4
\cr && +248 e_5+56 e_6)+\mathcal{O}(e_i^2)\,.\label{Tx}
\eea
Similarly, substituting $r_0=x\ell$ into (\ref{entropy-5D-1order}) provided $k=-1$ yields
\bea
S(x)&=&\fft{4\pi\omega_{-1,3}}{\ell_0 x}(x^4 (-2220 e_1-536 e_2-144 e_3-121 e_4-436 e_5-96 e_6+\ell_0^4)
\cr && +(4320 e_1+1056 e_2+288 e_3+240 e_4+864 e_5+192 e_6) x^2-1800 e_1
\cr && -480 e_2-144 e_3-114 e_4-408 e_5-96 e_6)+\mathcal{O}(e_i^2)\,.\label{Sx}
\eea
Notice in this section we would not divide out the possibly divergent volume $\omega_{-1,d-1}$ and refer the relevant quantity as the density, instead we shall keep in mind that $\omega_{-1,d-1}$ is the important ingredient encoding the universal piece of EE and R\'enyi entropy (see, e.g. \cite{Calabrese:2004eu,Ryu:2006bv}). From (\ref{Tx}) we can immediately know
\bea
&& T_0=\frac{3 \ell_0^4+60 e_1+8 e_2+e_4+4 e_5}{6 \pi  \ell_0^5}+\mathcal{O}(e_i^2)\,,
\cr && \tilde{r}=\ell_0-\fft{60 e_1+8 e_2+e_4+4 e_5}{3\ell_0^3}+\mathcal{O}(e_i^2)=\ell\,.\label{T0-r}
\eea
The second line in (\ref{T0-r}) implies the consistency of this procedure: In the boundary, the metric takes the form as
\be
ds_{\rm bound}^2=\fft{r^2}{\ell^2}(-dt^2+\ell^2d\Omega_{-1,3}^2)\,,
\ee
which is conformal to the hyperbolic cylinder $\mathbb{R}\times \mathbb{H}^{3}$ with the radius $\tilde{r}$
\be
ds_{\mathbb{R}\times \mathbb{H}^{3}}^2=-dt^2+\tilde{r}^2 d\Omega_{-1,3}^2\,.
\ee
The equation (\ref{xq}) is explicitly given by
\bea
&& x_q^6 (360 e_1 q+288 e_2 q+144 e_3 q+90 e_4 q+440 e_5 q+112 e_6 q-6 \ell_0^4 q)
\cr && +(60 e_1+8 e_2+e_4+4 e_5+3 \ell_0^4) x_q^5+x_q^4 (-300 e_1 q-400 e_2 q-216 e_3 q
\cr &&-131 e_4 q-644 e_5 q-168 e_6 q+3 \ell_0^4 q)+(-720 e_1 q-96 e_2 q-12 e_4 q-48 e_5 q) x_q^2
\cr && +600 e_1 q+200 e_2 q+72 e_3 q+52 e_4 q+248 e_5 q+56 e_6 q+\mathcal{O}(e_i^2)=0\,.\label{xq-eq0D=5}
\eea
Then substitute (\ref{Tx}), (\ref{Sx}), (\ref{T0-r}) and $x_q$ as the solution of (\ref{xq-eq0D=5}) into (\ref{Sq-formu}), we end up with the R\'enyi entropy
\bea
S_q &=& \frac{\pi\omega_{-1,3}}{4 \ell_0 (q-1) q^3 \sqrt{8 q^2+1} (\sqrt{8 q^2+1}+1)^2}  (6912 e_3 q^6+5536 e_4 q^6+13696 e_5 q^6+3840 e_6 q^6
\cr && -1440 e_3 q^4-1772 e_4 q^4-5328 e_5 q^4-800 e_6 q^4+288 e_3 q^2-8 e_4 \sqrt{8 q^2+1} q^2+292 e_4 q^2
\cr && -32 e_5 \sqrt{8 q^2+1} q^2+816 e_5 q^2+160 e_6 q^2+72 e_3 \sqrt{8 q^2+1}+75 e_4 \sqrt{8 q^2+1}
\cr && +212 e_5 \sqrt{8 q^2+1}
 +40 e_6 \sqrt{8 q^2+1}-1152 e_3 \sqrt{8 q^2+1} q^6-304 e_4 \sqrt{8 q^2+1} q^6
 \cr && +192 e_5 \sqrt{8 q^2+1} q^6
 -640 e_6 \sqrt{8 q^2+1} q^6-864 e_3 \sqrt{8 q^2+1} q^4-1140 e_4 \sqrt{8 q^2+1} q^4
 \cr && -3504 e_5 \sqrt{8 q^2+1} q^4
 -480 e_6 \sqrt{8 q^2+1} q^4+60 e_1 ((124-8 \sqrt{8 q^2+1}) q^2
 \cr && +33 (\sqrt{8 q^2+1}+1)
 +16 (23 \sqrt{8 q^2+1}+94) q^6-4 (159 \sqrt{8 q^2+1}+233) q^4)
 \cr && +64 e_2 (-q^2 (\sqrt{8 q^2+1}-23)
 +6 (\sqrt{8 q^2+1}+1)+16 (\sqrt{8 q^2+1}+23) q^6
 \cr && -2 (51 \sqrt{8 q^2+1}+77) q^4)+72 e_3+75 e_4
 +212 e_5+40 e_6+32 \ell_0^4 q^6-92 \ell_0^4 q^4-20 \ell_0^4 q^2
 \cr && -16 \ell_0^4 \sqrt{8 q^2+1} q^2-\ell_0^4 \sqrt{8 q^2+1}
 +80 \ell_0^4 \sqrt{8 q^2+1} q^6-36\ell_0^4 \sqrt{8 q^2+1} q^4-\ell_0^4)
 \cr && +\mathcal{O}(e_i)^2\,.\label{Renyi-D=5}
\eea
We then have access to the EE by taking $q\rightarrow 1$ in (\ref{Renyi-D=5}), we find
\be
S_{\rm EE}=\lim_{q\rightarrow 1}S_q=4\pi(\ell_0^3+5(60e_1+8e_2+e_4+4e_5)\ell_0^{-1})+\mathcal{O}(e_i)^2\,.
\ee
Note up to the first order, the $a$-charge in $D=5$, $d=4$ (\ref{a-5D}) is approximately given by
\be
a=2\pi^2(\ell_0^3+5(60e_1+8e_2+e_4+4e_5)\ell_0^{-1})+\mathcal{O}(e_i)^2\,,
\ee
therefore we have
\be
S_{\rm EE}=\fft{2 \omega_{-1,3}}{\pi} a\,.\label{EE-5D}
\ee
EE (\ref{EE-5D}) fits our expectation that EE should encode the $a$-charge in even $d$ dimensions \cite{Myers:2010xs,Myers:2010tj,Hung:2011xb,Myers:2012ed}. It is also of interest to investigate the limit $q\rightarrow 0$ where we obtain
\be
\lim_{q\rightarrow 0}S_q=\frac{\pi \omega_{-1,3}  (\ell_0^4-1980 e_1-384 e_2-72 e_3-75 e_4-212 e_5-40 e_6)}{8 \ell_0 q^3}+\mathcal{O}(e_i^2)\,.\label{D=5-q=0}
\ee
We then can verify the relation found in \cite{Hung:2011nu} for (\ref{D=5-q=0}) up to the first order of $e_i$
\be
\lim_{q\rightarrow 0}S_q=\fft{a\Big(3\fft{c}{a}(1+\fft{1}{630}t_4)-1\Big)^4}{\Big(5\fft{c}{a}(1+\fft{1}{945}t_4)-1\Big)^3}\fft{\omega_{-1,3}}{4\pi q^3}\,,
\ee
where $a$ is given in (\ref{a-5D}), $c$ is given in (\ref{c-5D}) and $t_2$, $t_4$ can be found in (\ref{t2-t4}). We then consider the large $q$ limit in which we have
\be
\lim_{q\rightarrow \infty}S_q=\frac{\pi \omega_{-1,3}  (5 \ell_0^4+1380 e_1+64 e_2-72 e_3-19 e_4+12 e_5-40 e_6)}{2 \ell_0}\,.\label{D=5-q=inf}
\ee
We can easily verify, up to the first order, (\ref{D=5-q=inf}) satisfies the relation found in \cite{Hung:2011nu}
\be
\lim_{q\rightarrow \infty}S_q=\fft{2\omega_{-1,3}}{\pi}a\Big(1+\fft{3}{2}\fft{(\fft{c}{a})^2}{1-5\fft{c}{a}}-\fft{t_4}{1935}\ft{1-17
\fft{c}{a}+98(\fft{c}{a})^2-194(\fft{c}{a})^3-17(\fft{c}{a})^4+215(\fft{c}{a})^5}{2\fft{c}{a}(1-3\fft{c}{a})(1-5\fft{c}{a})^2}\Big)\,.
\ee
In fact we find, up to the first order of $e_i$, $\lim_{q\rightarrow \infty}S_q$ can be controlled by $c/a$ and $t_4$ in various ways
\bea
 \lim_{q\rightarrow \infty}S_q=\fft{2\omega_{-1,3}}{\pi}a\Big(1+\fft{3}{2}\fft{(\fft{c}{a})^2}{1-5\fft{c}{a}}-\fft{t_4}{1935}\ft{1+c_1
\fft{c}{a}+c_2 (\fft{c}{a})^2+c_3 (\fft{c}{a})^3+c_4 (\fft{c}{a})^4+c_5 (\fft{c}{a})^5}{2\fft{c}{a}(1-3\fft{c}{a})(1-5\fft{c}{a})^2}\Big)\,,
\eea
where $c_i$s only need to satisfy $c_1+c_2+c_3+c_4+c_5=85$.
\subsubsection{$D=4$}
In $D=4$, the black holes are given in (\ref{f-k-D=4}) and (\ref{h-k-D=4}) provided with $k=-1$. Note the effective AdS radius $\ell$ is (\ref{1-ord-ell-4D}), therefore replacing $r_0=x\ell$ in (\ref{T-4D-1order}) leads to
\bea
T(x) &=& -\frac{1}{24 \pi  x^5 \ell _0^5}(x^6 (540 e_1+198 e_2+81 e_3+60 e_4+132 e_5+54 e_6-18 \ell _0^4)
\cr && +x^4 (-1548 e_1-474 e_2-162 e_3-133 e_4-316 e_5-108 e_6+6 \ell _0^4)
\cr && +(972 e_1+270 e_2+81 e_3+72 e_4+180 e_5+54 e_6) x^2-108 e_1-18 e_2
\cr && -3 e_4-12 e_5)+\mathcal{O}(e_i^2)\,.\label{Tx-D=4}
\eea
Similarly, from (\ref{entropy-4D-1order}) we have
\bea
S(x)&=&-\fft{2\pi\omega_{-1,2}}{3x^2\ell_0^2}(x^4 (1440 e_1+348 e_2+81 e_3+85 e_4+232 e_5+54 e_6-6 \ell _0^4)
\cr && +(-2592 e_1-648 e_2-162 e_3-162 e_4-432 e_5-108 e_6) x^2+864 e_1
\cr && +252 e_2+81 e_3+69 e_4+168 e_5+54 e_6)+\mathcal{O}(e_i^2)\,.\label{Sx-D=4}
\eea
We then have
\bea
&& T_0=\frac{3 \ell _0^4+36 e_1+6 e_2+e_4+4 e_5}{6 \pi  \ell _0^5}\,,
\cr && \tilde{r}=\ell_0-\fft{36 e_1+6 e_2+e_4+4 e_5}{3\ell_0^3}+\mathcal{O}(e_i^2)=\ell\,.\label{T0-r-D=4}
\eea
From the second line of (\ref{T0-r-D=4}), the consistency is manifest, the boundary metric is
\be
ds_{\rm bound}^2=\fft{r^2}{\ell^2}(-dt^2+\ell^2d\Omega_{-1,2}^2)\,,
\ee
which can be conformally mapped to the hyperbolic cylinder $\mathbb{R}\times \mathbb{H}^{2}$ with the radius $\tilde{r}$
\be
ds_{\mathbb{R}\times \mathbb{H}^{2}}^2=-dt^2+\tilde{r}^2 d\Omega_{-1,2}^2\,.
\ee
The equation (\ref{xq}) now is
\bea
&& x_q^6 (540 e_1 q+198 e_2 q+81 e_3 q+60 e_4 q+132 e_5 q+54 e_6 q-18 q \ell _0^4)+(144 e_1+24 e_2
\cr && +4 e_4+16 e_5+12 \ell _0^4) x_q^5+x_q^4 (-1548 e_1 q-474 e_2 q-162 e_3 q-133 e_4 q-316 e_5 q
\cr && -108 e_6 q+6 q \ell _0^4)+(972 e_1 q+270 e_2 q+81 e_3 q+72 e_4 q+180 e_5 q+54 e_6 q) x_q^2
\cr && -108 e_1 q-18 e_2 q-3 e_4 q-12 e_5 q+\mathcal{O}(e_i^2)=0\,.\label{xq-D=4}
\eea
Having (\ref{Tx-D=4}), (\ref{Sx-D=4}) and $x_q$ which can be solved by (\ref{xq-D=4}), we have the holographic R\'enyi entropy in $D=4$ for massless cubic gravities as
\bea
S_q&=&\frac{4\pi\omega_{-1,2}}{81 (q-1) q^2 \ell _0^2}\Big(\frac{2}{(\sqrt{3 q^2+1}+1)^3 (9 (\sqrt{3 q^2+1}+4) q^4+12 (2 \sqrt{3 q^2+1}+3) q^2+8 (\sqrt{3 q^2+1}+1))}
\cr && \big(4374 e_4 q^{11}+17496 e_5 q^{11}-6561 \sqrt{3 q^2+1} e_3 q^{10}-6561 e_3 q^{10}-4131 \sqrt{3 q^2+1} e_4 q^{10}
\cr && -2673 e_4 q^{10}-7776 \sqrt{3 q^2+1} e_5 q^{10}-1944 e_5 q^{10}-4374 \sqrt{3 q^2+1} e_6 q^{10}-4374 e_6 q^{10}
\cr && +10206 \sqrt{3 q^2+1} e_4 q^9+36450 e_4 q^9+40824 \sqrt{3 q^2+1} e_5 q^9+145800 e_5 q^9+8748 \sqrt{3 q^2+1} e_3 q^8
\cr && +21870 e_3 q^8+1134 \sqrt{3 q^2+1} e_4 q^8-18792 e_4 q^8-7128 \sqrt{3 q^2+1} e_5 q^8-104328 e_5 q^8
\cr && +5832 \sqrt{3 q^2+1} e_6 q^8+14580 e_6 q^8+27216 \sqrt{3 q^2+1} e_4 q^7+50544 e_4 q^7+108864 \sqrt{3 q^2+1} e_5 q^7
\cr && +202176 e_5 q^7+3645 \sqrt{3 q^2+1} e_3 q^6-9477 e_3 q^6-30753 \sqrt{3 q^2+1} e_4 q^6-71415 e_4 q^6
\cr && -127872 \sqrt{3 q^2+1} e_5 q^6-273024 e_5 q^6+2430 \sqrt{3 q^2+1} e_6 q^6-6318 e_6 q^6+18144 \sqrt{3 q^2+1} e_4 q^5
\cr && +23328 e_4 q^5+72576 \sqrt{3 q^2+1} e_5 q^5+93312 e_5 q^5-9720 \sqrt{3 q^2+1} e_3 q^4-9720 e_3 q^4
\cr && -29448 \sqrt{3 q^2+1} e_4 q^4-31176 e_4 q^4-104832 \sqrt{3 q^2+1} e_5 q^4-111744 e_5 q^4-6480 \sqrt{3 q^2+1} e_6 q^4
\cr && -6480 e_6 q^4+3456 \sqrt{3 q^2+1} e_4 q^3+3456 e_4 q^3+13824 \sqrt{3 q^2+1} e_5 q^3+13824 e_5 q^3
\cr && +1296 \sqrt{3 q^2+1} e_3 q^2+3888 e_3 q^2+816 \sqrt{3 q^2+1} e_4 q^2+4752 e_4 q^2+1536 \sqrt{3 q^2+1} e_5 q^2
\cr && +13824 e_5 q^2+864 \sqrt{3 q^2+1} e_6 q^2+2592 e_6 q^2-72 (-2187 q^{11}+243 (\sqrt{3 q^2+1}-2) q^{10}
\cr && -729 (7 \sqrt{3 q^2+1}+25) q^9+81 (23 \sqrt{3 q^2+1}+191) q^8-1944 (7 \sqrt{3 q^2+1}+13) q^7
\cr && +27 (607 \sqrt{3 q^2+1}+1225) q^6-1296 (7 \sqrt{3 q^2+1}+9) q^5+72 (167 \sqrt{3 q^2+1}+179) q^4
\cr && -1728 (\sqrt{3 q^2+1}+1) q^3-48 (\sqrt{3 q^2+1}+27) q^2-832 (\sqrt{3 q^2+1}+1)) e_1-12 (-2187 q^{11}
\cr && +243 (4 \sqrt{3 q^2+1}+1) q^{10}-729 (7 \sqrt{3 q^2+1}+25) q^9+81 (11 \sqrt{3 q^2+1}+161) q^8
\cr && -1944 (7 \sqrt{3 q^2+1}+13) q^7+432 (37 \sqrt{3 q^2+1}+79) q^6-1296 (7 \sqrt{3 q^2+1}+9) q^5
\cr && +144 (91 \sqrt{3 q^2+1}+97) q^4-1728 (\sqrt{3 q^2+1}+1) q^3-192 (\sqrt{3 q^2+1}+9) q^2
\cr && -1024 (\sqrt{3 q^2+1}+1)) e_2+1728 \sqrt{3 q^2+1} e_3+1728 e_3+2624 \sqrt{3 q^2+1} e_4+2624 e_4
\cr && +8192 \sqrt{3 q^2+1} e_5+8192 e_5+1152 \sqrt{3 q^2+1} e_6+1152 e_6\big)-3 (-27 q^3
\cr && +(6 \sqrt{3 q^2+1}+9) q^2+2 (\sqrt{3 q^2+1}+1)) \ell _0^4\Big)+\mathcal{O}(e_i)^2\,.\label{Renyi-D=4}
\eea
Although (\ref{Renyi-D=4}) looks cumbersome and ugly, its limit $q\rightarrow 1$ gives satisfactory value
\be
S_{\rm EE}=\lim_{q\rightarrow 1}S_q=\fft{4\pi\omega_{-1,2}}{3}(3\ell_0^2+4(36e_1+6e_2+e_4+4e_5)\ell_0^{-2})+\mathcal{O}(e_i)^2\,.
\ee
Note the $a$-charge defined in general dimensions (\ref{a-charge-d}) specialized in $d=3$ gives rise to
\be
a=4\pi(\ell^2+2(36e_1+6e_2+e_4+4e_5)\ell^{-2})=4\pi(\ell_0^2+\fft{4}{3}(36e_1+6e_2+e_4+4e_5)\ell_0^{-2})+\mathcal{O}(e_i)^2\,,
\ee
we then conclude
\be
S_{\rm EE}=\omega_{-1,2}\, a\,.\label{SEE-4D}
\ee
Actually, (\ref{SEE-4D}) serves as one of the reason that the $a$-charge defined in odd $d$ dimensions is also meaningful \cite{Myers:2010xs,Myers:2010tj,Hung:2011xb,Myers:2012ed}. Furthermore, we have $q\rightarrow 0$ limit behaving
\bea
\lim_{q\rightarrow 0}S_q=-\frac{8 \pi  \left(936 e_1+192 e_2+27 e_3+41 e_4+128 e_5+18 e_6-6 \ell _0^4\right)}{81 q^2 \ell _0^2}+\mathcal{O}(e_i)^2\,,\label{D=4-Sq-q=0}
\eea
and large $q$ limit result is given by
\bea
\lim_{q\rightarrow \infty}S_q &=& -\frac{4 \pi}{27 \ell _0^2}  (2 (72 (\sqrt{3}-9) e_1+12 (4 \sqrt{3}-9) e_2+27 \sqrt{3} e_3+17 \sqrt{3} e_4-18 e_4
\cr && +32 \sqrt{3} e_5-72 e_5+18 \sqrt{3} e_6)+3 (2 \sqrt{3}-9) \ell _0^4)+\mathcal{O}(e_i)^2\,.\label{D=4-Sq-q=inf}
\eea
\subsection{Energy flux parameters from twist operators}
In the context of the replica trick, $q$ copies of the background geometry should be glued together as the $q$-fold manifold with some "twist" boundary condition \cite{Calabrese:2004eu}, in which the boundary conditions can be implemented by inserting the twist operators $\sigma_q$ \cite{Hung:2011nu,Hung:2014npa,Swingle:2010jz,Calabrese:2004eu}. The scaling dimension of the twist operators $h_q$ actually encodes the information of $t_2$ and $t_4$ parameters for a CFT \cite{Chu:2016tps}. It turns out that specifically we have \cite{Chu:2016tps}
\bea
\fft{h_q''(q=1)}{C_T} &=& -\fft{2\pi^{1+\fft{d}{2}}\Gamma(\fft{d}{2})}{(d-1)^3 d(d+1)\Gamma(d+3)}\Big(d(2d^5-9d^3+2d^2+7d-2)
\cr &&
\cr &&
 +(d-2)(d-3)
(d+1)(d+2)(2d-1)t_2
 +(d-2)(7d^3+9d^2-8d+8)t_4\Big)\,,
 \cr &&\label{hq-t2-t4}
\eea
where the prime stands for the derivatives with respect to $q$. On the other hand, in the thermodynamics viewpoint of EE and R\'enyi entropy in $\mathbb{R}\times\mathbb{H}^{d-1}$, the scaling dimension $h_q$ can be calculated directly by \cite{Hung:2011nu,Hung:2014npa}
\be
h_q=\fft{2\pi\tilde{r}q}{(d-1)\omega_{-1,d-1}}\int_{x_q}^1 S'(x)T(x)dx\,.\label{hq}
\ee
Therefore, in the holographic context, as one readily obtains $h_q$ as a function of $q$ from (\ref{hq}), one can immediately make use of (\ref{hq-t2-t4}) to verify the results of $t_2$ and $t_4$. More surprisingly, in $d=3$ where the holographic energy flux method adopted in section \ref{flux} is not convenient, the formula (\ref{hq-t2-t4}) can be viewed as a powerful tool to determine $t_4$ \cite{Bueno:2018xqc}.

In this subsection, we would obtain $h_q$ by using (\ref{hq}) for $D=5$ and $D=4$ approximate hyperbolic black holes in massless cubic gravities respectively, and then make use of (\ref{hq-t2-t4}) to verify $t_2$ and $t_4$ (\ref{t2-t4}) obtained in section \ref{flux} for $D=5$. Most importantly, (\ref{hq-t2-t4}) shall be employed to obtain $t_4$ approximately up to the first order of $e_i$ for $D=4$, then the approximate result shall be enhanced to be the exact one.
\subsubsection{$D=5$}
For $D=5$ black holes (\ref{f-k-D=5}) and (\ref{h-k-D=5}) with $k=-1$, Substitute (\ref{Tx}), (\ref{Sx}) and (\ref{T0-r}) into (\ref{hq}), we have
\bea
\fft{h_q}{C_T}&=&\frac{\pi ^3}{960 q^3 \sqrt{8 q^2+1} (\sqrt{8 q^2+1}+1)^2 \ell _0^4} (8 (288 e_3 q^6+272 e_4 q^6+736 e_5 q^6+160 e_6 q^6-252 e_3 q^4
\cr && -256 e_4 q^4-716 e_5 q^4-140 e_6 q^4+72 e_3 \sqrt{8 q^2+1} q^2+180 e_3 q^2+65 e_4 \sqrt{8 q^2+1} q^2+173 e_4 q^2
\cr && +172 e_5 \sqrt{8 q^2+1} q^2+472 e_5 q^2+40 e_6 \sqrt{8 q^2+1} q^2+100 e_6 q^2+27 e_3 \sqrt{8 q^2+1}
\cr && +27 e_4 \sqrt{8 q^2+1}+75 e_5 \sqrt{8 q^2+1}+15 e_6 \sqrt{8 q^2+1}+144 e_3 \sqrt{8 q^2+1} q^6+136 e_4 \sqrt{8 q^2+1} q^6
\cr && +368 e_5 \sqrt{8 q^2+1} q^6+80 e_6 \sqrt{8 q^2+1} q^6-324 e_3 \sqrt{8 q^2+1} q^4-300 e_4 \sqrt{8 q^2+1} q^4
\cr && -804 e_5 \sqrt{8 q^2+1} q^4-180 e_6 \sqrt{8 q^2+1} q^4+15 e_1 (4 (23 \sqrt{8 q^2+1}+68) q^2+45 (\sqrt{8 q^2+1}+1)
\cr && +208 (\sqrt{8 q^2+1}+2) q^6-4 (111 \sqrt{8 q^2+1}+109) q^4)+e_2 (4 (76 \sqrt{8 q^2+1}+211) q^2
\cr && +135 (\sqrt{8 q^2+1}+1)+656 (\sqrt{8 q^2+1}+2) q^6-68 (21 \sqrt{8 q^2+1}+19) q^4)+27 e_3+27 e_4
\cr && +75 e_5+15 e_6)+3 (-4 (2 \sqrt{8 q^2+1}+3) q^2-\sqrt{8 q^2+1}+16 (\sqrt{8 q^2+1}+2) q^6
\cr && -4 (\sqrt{8 q^2+1}+7) q^4-1) \ell _0^4)+\mathcal{O}(e_i^2)\,.\label{hq-D=5}
\eea
Subsequently we have
\be
\fft{h_q''(q=1)}{C_T}=-\frac{\pi ^3 \Big(17 \ell _0^4-8 (1620 e_1+336 e_2+72 e_3+69 e_4+188 e_5+40 e_6)\Big)}{540 \ell _0^4}+\mathcal{O}(e_i^2)\,.\label{hqpp}
\ee
It is easy to verify that substituting (\ref{t2-t4}) into the right hand of (\ref{hq-t2-t4}) and expanding it up to the linear order of $e_i$ immediately recover (\ref{hqpp}).

One can also immediately verify some other identities up to the first order of $e_i$ \cite{Bueno:2015qya,Hung:2014npa}\footnote{Note our convention is a little different from \cite{Bueno:2015qya,Hung:2014npa,Bueno:2018xqc} by an overall $\tilde{r}^{d-1}=\ell^{d-1}$ factor in the definition of the entropy $S$, which is also consistent.}
\bea
&& h_q'(q=1)=\fft{2\pi^{\fft{d}{2}+1}\Gamma(\fft{d}{2})}{\Gamma(d+2)}C_T\,,\qquad \lim_{q\rightarrow 0}h_q=-\fft{C_s}{d\,\tilde{r}^{d-1}}\Big(
\fft{1}{2\pi q}\Big)^{d-1}\,,
\cr &&
\cr && \partial_q^j h_q(q=1)=-\fft{1}{(d-1)\omega_{-1,d-1}}\Big((j+1)\partial_q^j S_q+j^2 \partial_q^{j-1}S_q\Big)\Big|_{q=1}\,,\label{identities}
\eea
where the second term in the last line shall be dropped for $j=1$, and $C_s$ is defined for planar black holes
\be
C_s=\fft{s}{T^{d-1}}\,,
\ee
where $s$ is the entropy density and $T$ is the temperature. For $D=5$ approximate planar black holes we have, up to the leading order
\be
C_s=4 \pi ^4 \ell _0^2 (\ell _0^4-4 (510 e_1+98 e_2+18 e_3+19 e_4+54 e_5+10 e_6))+\mathcal{O}(e_i^2)\,.
\ee
\subsubsection{$D=4$}
For $D=4$ black holes (\ref{f-k-D=4}) and (\ref{h-k-D=4}) with $k=-1$, substituting (\ref{Tx-D=4}), (\ref{Sx-D=4}) and (\ref{T0-r-D=4}) into (\ref{hq}) yields
\bea
\fft{h_q}{C_T} &=& \frac{\pi ^3}{972 q^2 (\sqrt{3 q^2+1}+1)^3 (12 (2 \sqrt{3 q^2+1}+3) q^2+8 (\sqrt{3 q^2+1}+1)+9 (\sqrt{3 q^2+1}+4) q^4) \ell _0^4}
\cr && ((360 e_1+96 e_2+27 e_3+25 e_4+64 e_5+18 e_6) (96 (7 \sqrt{3 q^2+1}+9) q^2+128 (\sqrt{3 q^2+1}+1)
\cr && +243 (\sqrt{3 q^2+1}+7) q^{10}+81 (11 \sqrt{3 q^2+1}-7) q^8-54 (31 \sqrt{3 q^2+1}+49) q^6
\cr && -288 (\sqrt{3 q^2+1}-2) q^4)+3 (-192 (7 \sqrt{3 q^2+1}+9) q^2-256 (\sqrt{3 q^2+1}+1)
\cr && +243 (\sqrt{3 q^2+1}+7) q^{10}+81 (23 \sqrt{3 q^2+1}+47) q^8+216 (5 \sqrt{3 q^2+1}-1) q^6
\cr && -144 (11 \sqrt{3 q^2+1}+23) q^4) \ell _0^4)+\mathcal{O}(e_i^2)\,.
\eea
We still have (\ref{identities}) with $C_s$ now given by
\be
C_s=\frac{32}{9} \pi ^3 (2 \ell_0^4-3 (120 e_1+24 e_2+3 e_3+5 e_4+16 e_5+2 e_6))+\mathcal{O}(e_i^2)\,.
\ee
Then we also have
\be
\fft{h_q''(q=1)}{C_T}=\frac{\pi ^3 (720 e_1+192 e_2+54 e_3+50 e_4+128 e_5+36 e_6-7 \ell _0^4)}{96 \ell _0^4}+\mathcal{O}(e_i^2)\,.\label{hqq-D=4}
\ee
Compare the result in $D=4$ (\ref{hqq-D=4}) with (\ref{hq-t2-t4}) specialized in $d=3$, we can solve $t_4$ for the massless cubic gravities in $D=4$ up to the first order of $e_i$
\be
t_4=-\frac{120 (360 e_1+96 e_2+27 e_3+25 e_4+64 e_5+18 e_6)}{\ell _0^4}+\mathcal{O}(e_i^2)\,.\label{t4-D=4-1order}
\ee
In fact, recall the original definition of the energy flux (\ref{energy-flux-defin}), it is obvious that the denominator of $t_4$ is exclusively proportional to the two-point function coefficient $C_T$, implying that one can simply enhance the first order result (\ref{t4-D=4-1order}) of $t_4$ to be a non-perturbative result
\be
t_4=\frac{120 (360 e_1+96 e_2+27 e_3+25 e_4+64 e_5+18 e_6)}{-\ell ^4+72 e_1+12 e_2+2 e_4+8 e_5}\,,\label{t4-D=4}
\ee
where $\ell$ is the effective AdS radius. For Einsteinian cubic gravity in $D=4$ where the coupling constants are given as (we follow the notations in \cite{Bueno:2018xqc})
\bea
(e_1,e_2,e_3,e_4,e_5,e_6,e_7,e_8)=-\fft{\mu\ell_0^4}{8}\,\Big(0,0,8,-12,0,0,1,-12\Big)\,,
\label{Einsteinian-cubic}
\eea
after introducing $f_{\infty}=\ell_0^2/\ell^2$, (\ref{t4-D=4}) would match the exact result obtained in \cite{Bueno:2018xqc} (without any perturbative treatment)
\be
t_4=-\fft{1260f_{\infty}^2\mu}{1-3f_{\infty}^2 \mu}\,.
\ee

Provided $t_4$ in $D=4$ (\ref{t4-D=4}), we find there are several ways to reexpress $\lim_{q\rightarrow0}S_q$ (\ref{D=4-Sq-q=0}) in terms of $\tilde{a}$, $\mathcal{C}_T$ and $t_4$ up to the first order, for example, we find
\be
\lim_{q\rightarrow 0}S_q=\fft{\tilde{a}\Big(4\fft{\mathcal{C}_T}{\tilde{a}}(1+\fft{1}{1800}t_4)-1\Big)^4}{\Big(3\fft{\mathcal{C}_T}{\tilde{a}}
(1+\fft{1}{3600}t_4)-1\Big)^3}\fft{64\pi \omega_{-1,3}}{729 q^3}\,,
\ee
where $\tilde{a}$, $\mathcal{C}_T$ are given in the first line and the third line in (\ref{a-c-CT}) provided $d=3$
\be
\tilde{a}=\ell^2+2(36e_1+6e_2+e_4+4e_5)\ell^{-2}\,,\qquad \mathcal{C}_T=\ell^2-2(36e_1+6e_2+e_4+4e_5)\ell^{-2}\,.\label{a-CT-d=3}
\ee
Up to the first order of $e_i$, we can even find some ways to use $\mathcal{C}_T/\tilde{a}$ and $t_4$ to control the behavior of $\lim_{q\rightarrow\infty}S_q$ (\ref{D=4-Sq-q=inf}), for instance
\bea
\lim_{q\rightarrow \infty}S_q &=& \ft{92(2\sqrt{3}-9)\omega_{-1,2}}{9(12\sqrt{3}-31)}\tilde{a}\Big(1+\ft{8(3\sqrt{3}-2)}{23}
\fft{(\fft{\mathcal{C}_T}{\tilde{a}})^2}{1-3\fft{\mathcal{C}_T}{\tilde{a}}}
\cr && +\ft{31\sqrt{3}-36}{230(2\sqrt{3}-9)} \ft{t_4}{\fft{\mathcal{C}_T}{\tilde{a}}(1-3\fft{\mathcal{C}_T}{\tilde{a}})(1-4\fft{\mathcal{C}_T}{\tilde{a}})^2}\Big)\,.
\eea

In the last, we turn to present the physical constraints in $D=4$ (\ref{D=4-constriant}), from $\mathcal{C}_T>0$
\be
36e_1+6e_2+e_4+4e_5<\fft{\ell^4}{2}\,,\label{c-cons-D=4}
\ee
and from $\langle\mathcal{E}\rangle\geq0$
\be
\Big|5436 e_1+1446 e_2+405 e_3+376 e_4+964 e_5+270 e_6\Big|\leq\fft{\ell^4}{2}\,.\label{t4-cons-D=4}
\ee
Again, treating $e_i$ infinitesimal compared to any other theory constants is safe enough to satisfy the constraints (\ref{c-cons-D=4}) and (\ref{t4-cons-D=4}).
\section{Hydrodynamics}
\label{hydro}
In this section, we study the holographic hydrodynamics, more specifically, the shear-viscosity-entropy-ratio for the massless cubic gravities. The shear-viscosity-entropy-ratio is a very important transport property of holographic hydrodynamics and it was studied considerably in the literature both for Einstein gravity extended with higher order corrections and gravity theories coupled with matter fields \cite{Myers:2010jv,Buchel:2009sk,Banerjee:2009fm,deBoer:2009gx,Li:2017ncu,Peng:2018vbe,Bueno:2018xqc,Kats:2007mq,Policastro:2001yc,KSS,KSS0,Buchel:2003tz,Buchel:2004qq,Benincasa:2006fu,
Landsteiner:2007bd,Cremonini:2011iq,Iqbal:2008by,Cai:2008ph,Cai:2009zv,Brustein:2007jj,
Liu:2015tqa,Liu:2016way,Hartnoll:2016tri,Alberte:2016xja,Liu:2016njg,Wang:2016vmm,Ge:2015owa,Ge:2014aza,Paulos:2009yk} . In this section, we use the ``pole method'' proposed in \cite{Paulos:2009yk} (see also e.g. \cite{Myers:2010jv,Bueno:2018xqc}) to calculate the shear-viscosity-entropy-ratio for the second order approximate planar black holes in $D=5$ and $D=4$ respectively of the massless cubic gravities where the black holes solutions are presented in Appendix \ref{2order}, (\ref{f-k=0-D=5-2order}) and (\ref{h-k=0-D=5-2order}) for $D=5$ and (\ref{f-k=0-D=4-2order}) and (\ref{h-k=0-D=4-2order}) for $D=4$. Consequently, the shear-viscosity-entropy-ratio is also expanded up to the second order of $e_i$ which can provide the information about how it can deviate from KSS bound \cite{KSS,KSS0} under the effect of the higher order corrections. Afterwards, we would like to try to express the deviations from KSS bound in terms of $(c-a)/c$ and $t_4$ (in $D=4$, we choose $(\mathcal{C}_T-\tilde{a})/\mathcal{C}_T$) as in, e.g. \cite{Banerjee:2009fm,Kats:2007mq}.

\subsection{$D=5$}
We start with the black hole background (\ref{bh}) in $D=5$ where $f$ and $h$ are given in (\ref{f-k=0-D=5-2order}) and (\ref{h-k=0-D=5-2order}) respectively. Then we impose the off-shell perturbation as follows
\be
dx^1\rightarrow dx^1+\varepsilon e^{-{\rm i}\omega t}dx^2\,.\label{off-shell-pert}
\ee
Note even in the metric, $\varepsilon$ should be kept in the second order. Afterwards, we substitute (\ref{off-shell-pert}) into the bulk Lagrangian $L$ in (\ref{bulk-cubic}) and expand it with respect to $\varepsilon$ up to the second order. The off-shell perturbation would create singular poles that are located at the horizon $r_0$. The ``pole method'' states that the shear-viscosity can be obtained by using the following formula
\be
\eta=-8\pi T \lim_{\omega\rightarrow 0,\varepsilon\rightarrow 0} \fft{{\rm Re}_{r=r_0}L }{\omega^2 \varepsilon^2}\,,\label{pole}
\ee
where $T$ is the temperature of the black hole (\ref{f-k=0-D=5-2order}) and (\ref{h-k=0-D=5-2order}), and $T$ is explicitly given by (\ref{T-order2-D=5}). Using (\ref{pole}) for (\ref{f-k=0-D=5-2order}) and (\ref{h-k=0-D=5-2order}), we have
\bea
\eta &=& \frac{r_0^3}{3 \ell _0^8} (48 (1095 e_1+251 e_2+63 e_3+55 e_4+135 e_5+34 e_6) \ell _0^4-128 (8433900 e_1^2
\cr && +60 (68354 e_2+18522 e_3+15544 e_4+47874 e_5+11413 e_6) e_1+499196 e_2^2+36828 e_3^2
\cr && +25859 e_4^2+223308 e_5^2+13388 e_6^2+61704 e_3 e_4+184824 e_3 e_5+156336 e_4 e_5+44748 e_3 e_6
\cr && +37660 e_4 e_6+109548 e_5 e_6+4 e_2 (67734 e_3+56798 e_4+172758 e_5+41471 e_6))
\cr && +3 \ell _0^8)+\mathcal{O}(e_i^3)\,.
\eea
Dividing by the entropy density (\ref{s-order2-D=5}) leads to the result
\bea
\fft{\eta}{s}&=&\fft{1}{4\pi}\Big(1+\frac{8}{\ell _0^4} (2460 e_1+568 e_2+144 e_3+125 e_4+324 e_5+80 e_6)-\frac{64}{3 \ell _0^8} (15831000 e_1^2
\cr && +30 (259820 e_2+71460 e_3+59275 e_4+187392 e_5+44404 e_6) e_1+960640 e_2^2
\cr && +73008 e_3^2+50083 e_4^2+463464 e_5^2+27160 e_6^2+120906 e_3 e_4+373320 e_3 e_5
\cr && +311466 e_4 e_5+89640 e_3 e_6+74492 e_4 e_6+224568 e_5 e_6+2 e_2 (264600 e_3+219305 e_4
\cr && +685884 e_5+163508 e_6))\Big)+\mathcal{O}(e_i^3)\,.\label{ratio-D=5}
\eea
For Myers quasi-topological gravity (\ref{Myers-quasi}) \cite{Myers:2010ru,Myers:2010jv}, (\ref{ratio-D=5}) reduces to be
\be
\fft{\eta}{s}=\fft{1}{4\pi}(1-324\mu-1728\mu^2)\,,
\ee
which is exactly the same as one obtained in \cite{Myers:2010jv}.

Note the first order result of the shear-viscosity-entropy-ratio is as follows
\bea
\fft{\eta}{s}=\fft{1}{4\pi}\Big(1+\frac{8}{\ell _0^4} (2460 e_1+568 e_2+144 e_3+125 e_4+324 e_5+80 e_6)\Big)+\mathcal{O}(e_i^2)\,.\label{ratio-D=5-order1}
\eea
We find, (\ref{ratio-D=5-order1}) can be uniquely expressed in terms of $(c-a)/c$ and $t_4$
\be
\fft{\eta}{s}=\fft{1}{4\pi}\Big(1-\fft{c-a}{c}-\fft{4}{45}t_4\Big)+\mathcal{O}(e_i)\,,\label{ratio-a-c-t}
\ee
where $a$ and $c$ are given in (\ref{a-5D}) and (\ref{c-5D}) respectively and $t_4$ can be found in (\ref{t2-t4}). (\ref{ratio-a-c-t}) is a surprise: we have $6$ coupling constants in (\ref{ratio-D=5-order1}) which should be expressed in terms of $5$ independent combinations\footnote{We do not count the Lovelock combination which is trivial in $D=5$.}, however, three algebraic independent combinations $a$, $c$ and $t_4$ are surprisingly enough to express (\ref{ratio-D=5-order1}). Therefore, in $D=5$ massless cubic gravities, the first order deviation from the KSS bound $1/(4\pi)$ can be totally controlled by the universal parameters of the corresponding CFT $(c-a)/c$ and $t_4$. Unfortunately, for the second order result (\ref{ratio-D=5}), we find $(c-a)/c$, $t_4$ and even $t_2$ may be not enough to determine the deviation, implying the linearized quasi-topological condition $a=c$ (see \cite{Li:2017txk} for more details about this condition) and the condition $t_4=0$ is not safe enough to guarantee $\eta/s$ saturates the KSS bound $1/(4\pi)$ up to the second order. Explicitly, the linearized quasi-topological condition $a=c$ imposes a constraint for coupling constants \cite{Li:2017txk}
\be
60e_1+8e_2+e_4+4e_5=0\,.\label{a=c}
\ee
If and only if the condition (\ref{a=c}) is satisfied, the perturbative treatment throughout this paper would not be necessary \cite{Li:2017txk}. For linearized quasi-topological cubic gravity (i.e. the constraint (\ref{a=c}) is satisfied), we have
\bea
\fft{\eta}{s} &=& \fft{1}{4\pi}\Big(1-\frac{32}{\ell _0^4} (600 e_1+20 e_2-36 e_3-11 e_4-20 e_6)-\frac{64}{3 \ell _0^8} (35784000 e_1^2+1200 (3152 e_2
\cr && -2880 e_3-686 e_4-1697 e_6) e_1+70960 e_2^2+73008 e_3^2+1183 e_4^2+27160 e_6^2+27576 e_3 e_4
\cr && +89640 e_3 e_6+18350 e_4 e_6-40 e_2 (5436 e_3+1595 e_4+3053 e_6))\Big)+\mathcal{O}(e_i^3)\,.
\eea
In addition, it turns out that Lovelock gravities \cite{Buchel:2009sk,deBoer:2009gx,Camanho:2013pda} and supersymmetric theories \cite{Hofman:2008ar,Kulaxizi:2009pz} should have $t_4=0$, hence $t_4=0$ might serve as additional important condition for gravity theories such that gravity theories can be more like Lovelock gravities, or can admit the potential for being enhanced to be supergravities. The condition $t_4=0$ together with $a=c$ (\ref{a=c}) requires
\bea
 600e_1+20e_2-36e_3-11e_4-20e_6=0\,.\label{t4=0-D=5}
\eea
It can be verified that (\ref{a=c}) together with (\ref{t4=0-D=5}) implies $t_2=0$. Imposing (\ref{a=c}) and (\ref{t4=0-D=5}) simultaneously, we have
\be
\fft{\eta}{s}=\fft{1}{4\pi}\Big(1+\fft{256}{5\ell_0^8}(600e_1+100e_2+12e_3+17e_4)^2\Big)+\mathcal{O}(e_i^3)\,.\label{ratio-D=5-free}
\ee

\subsection{$D=4$}
For $D=4$ black holes (\ref{f-k=0-D=4-2order}) and (\ref{h-k=0-D=4-2order}), follow the same procedure (note the temperature is given in (\ref{T-order2-D=4})) in the previous subsection, we have the shear-viscosity taking the form as
\bea
\eta &=& \fft{r_0^2}{4\ell_0^8}\Big(18 (672 e_1+180 e_2+51 e_3+47 e_4+120 e_5+34 e_6) \ell _0^4-27 (1267200 e_1^2
\cr && +24 (27732 e_2+7599 e_3+7179 e_4+18296 e_5+5066 e_6) e_1+88128 e_2^2+6975 e_3^2
\cr && +5981 e_4^2+38400 e_5^2+3100 e_6^2+12852 e_3 e_4+32376 e_3 e_5+30280 e_4 e_5+9300 e_3 e_6
\cr && +8568 e_4 e_6+21584 e_5 e_6+12 e_2 (4083 e_3+3821 e_4+9696 e_5+2722 e_6))+4 \ell _0^8\Big)
\cr && +\mathcal{O}(e_i^3)\,.
\eea
Dividing by the entropy density leads to the shear-viscosity-entropy-ratio
\bea
\fft{\eta}{s} &=& \fft{1}{4\pi}\Big(1+\frac{9}{\ell _0^4} (360 e_1+96 e_2+27 e_3+25 e_4+64 e_5+18 e_6)-\frac{27}{4 \ell _0^8} (1209600 e_1^2
\cr && +24 (26652 e_2+7389 e_3+6905 e_4+17768 e_5+4926 e_6) e_1+85248 e_2^2+6885 e_3^2
\cr && +5795 e_4^2+37888 e_5^2+3060 e_6^2+12576 e_3 e_4+31944 e_3 e_5+29592 e_4 e_5+9180 e_3 e_6
\cr && +8384 e_4 e_6+21296 e_5 e_6+12 e_2 (3993 e_3+3699 e_4+9472 e_5+2662 e_6))\Big)
\cr && +\mathcal{O}(e_i^3)\,.\label{ratio-D=4}
\eea
For $D=4$ Einsteinian cubic gravity (\ref{Einsteinian-cubic}), (\ref{ratio-D=4}) becomes
\be
\fft{\eta}{s}=\fft{1}{4\pi}\Big(1+\fft{189}{2}\mu-\fft{114453}{16}\mu^2\Big)\,,
\ee
which is the same as the result in \cite{Bueno:2018xqc}. Strikingly, (\ref{ratio-D=4}) can be uniquely controlled by $(\mathcal{C}_T-\tilde{a})/\mathcal{C}_T$ and $t_4$
\be
\fft{\eta}{s}=\fft{1}{4\pi}\Big(1-\fft{3}{40}t_4-\fft{45}{2}\Big(\fft{\mathcal{C}_T-\tilde{a}}{\mathcal{C}_T}\Big)^2
-\fft{17}{3840}t_4^2+\fft{9}{320}t_4 \Big(\fft{\mathcal{C}_T-\tilde{a}}{\mathcal{C}_T}\Big) \Big)\,,
\ee
where $\tilde{a}$ and $\mathcal{C}_T$ are given in (\ref{a-CT-d=3}).
The linearized quasi-topological condition is
\be
36e_1+6e_2+e_4+4e_5=0\,.\label{a=c-D=4}
\ee
We then have the shear-viscosity-entropy-ratio for the linearized quasi-topological gravity as
\bea
\fft{\eta}{s}&=&\fft{1}{4\pi}\Big(1-\frac{81}{\ell _0^4} (24 e_1-3 e_3-e_4-2 e_6)-\frac{20655}{4 \ell _0^8} (-24 e_1+3 e_3+e_4+2 e_6)^2\Big)
\cr && +\mathcal{O}(e_i^3)\,.
\eea
The requirement of $t_4$ implies
\be
24 e_1-3 e_3-e_4-2 e_6=0\,.
\ee
Hence, in $D=4$, the deviation from the KSS bound $1/(4\pi)$ provided $a=c$ and $t_4=0$ vanishes identically up to the second order.

\section{Conclusion}
\label{conclusion}
In this paper, we studied the holographic aspects of the generic massless cubic gravities coupled to a negative bare cosmological constant. In general, cubic gravities have $8$ combinations of higher-order curvature polynomials, while the decoupling of the massive spin-$2$ mode and massive scalar mode imposes two linear constraints such that the resulting gravity theories have $6$ coupling constants and they are called the massless cubic gravities. We focused on the discussions in $D=5$, $d=4$ and $D=4$, $d=3$, then we intended to complete the holographic dictionary for such generic massless cubic gravities. The holographic $a$-charge and the coefficient $C_T$-charge (and of course, the holographic $c$-charge in $D=5$ which is equivalent to $C_T$) appearing in the energy-momentum tensor two-point function were given in general dimensions in the literature. Then, to establish the holographic dictionary for the massless cubic gravities, the three-point function parameters $\mathcal{A}$, $\mathcal{B}$ and $\mathcal{C}$ or equivalently the energy flux parameters $t_2$ and $t_4$ should be obtained. Afterwards, treating the coupling constants as infinitesimal quantities, the generic massless cubic gravity theory would serve as an interesting hologrpahic model with adequate higher-order coupling constants to investigate how the effect of those higher-order coupling constants on some other CFT properties such as R\'enyi entropy and shear-viscosity-entropy-ratio can be controlled by the universal CFT parameters $c$, $a$ and $t_4$ (in $D=4$, there is $\mathcal{C}_T$, $\tilde{a}$ and $t_4$) perturbatively.

We obtained the boundary actions involving both the surface term and the holographic counterterms for the massless cubic gravities, and perturbatively, we solved out the approximate black holes expanded with the coupling constants $e_i$ by treating $e_i$s as very small quantities. Then, we analyzed the black hole thermodynamics with presenting important thermodynamic quantities such as the temperature and entropy which are useful for our purpose throughout this paper. Then, we calculated the energy flux parameters $t_2$ and $t_4$ in $D=5$ by considering the conformal collider thought experiment. In $D=4$, the situation becomes subtler and instead we obtained $t_4$ from the scaling dimension $h_q$ of the twist operators. With knowing $C_T$, $a$ and $t_2$, $t_4$ (recall in $D=4$, $t_2$ does not exist), the holographic dictionary was established. Taking the right coupling constants, the results in this paper nicely coincide with Myers quasi-topological gravity in $D=5$ and Einsteinian cubic gravity $D=4$. In $D=5$, $a$, $c$, $t_2$ and $t_4$ are listed in Table 1 with comparing to Myers quasi-topological gravity, and in $D=4$, $\tilde{a}$, $\mathcal{C}_T$, and $t_4$ are listed in Table 2 with comparing to Einsteinian cubic gravity.

\bigskip
\begin{table}[ht]
\begin{center}
\renewcommand\arraystretch{1.7}
\begin{tabular}{|c|c|c|}
  \hline
parameters & generic & Myers quasi-topological   \\ \hline

  $a$ & $2\pi^2(\ell^3 +6(60 e_1 + 8 e_2 + e_4 + 4 e_5)\ell^{-1})$ & $2\pi^2\ell_0^3 f_{\infty}^{-3/2}(1-3\mu f_{\infty}^2)$ \\

 \hline
$c$ & $2\pi^2(\ell^3 - 2(60 e_1 + 8 e_2 + e_4 + 4 e_5)\ell^{-1})$ & $2\pi^2\ell_0^3 f_{\infty}^{-3/2}(1+9\mu f_{\infty}^2)$\\
  \hline

$t_2$ & $\fft{48(2340e_1+552e_2+144e_3+123e_4+316e_5+80e_6)}{\ell^4-2(60e_1+8e_2+e_4+4e_5)}$ & $-\fft{2088\mu f_{\infty}^2}{1-3\mu f_{\infty}^2}$\\
  \hline
$t_4$ & $-\fft{360(600e_1+140e_2+36e_3+31e_4+80e_5+20e_6)}{\ell^4-2(60e_1+8e_2+e_4+4e_5)}$ & $\fft{3780\mu f_{\infty}^2}{1-3\mu f_{\infty}^2}$\\
  \hline
\end{tabular}
\caption{\small \it  The parameters $a$, $c$, $t_2$ and $t_4$ for the generic massless cubic gravities and Myers quasi-topological gravity in $D=5$. }
\end{center}
\end{table}

\bigskip
\begin{table}[ht]
\begin{center}
\renewcommand\arraystretch{1.7}
\begin{tabular}{|c|c|c|}
  \hline
parameters & generic & Einsteinian cubic gravity   \\ \hline

  $\tilde{a}$ & $\ell^2+2(36e_1+6e_2+e_4+4e_5)\ell^{-2}$ & $\ell_0^2 f_{\infty}^{-1}(1+3\mu f_{\infty}^2)$ \\

 \hline
$\mathcal{C}_T$ & $\ell^2-2(36e_1+6e_2+e_4+4e_5)\ell^{-2}$ & $\ell_0^2 f_{\infty}^{-1}(1-3\mu f_{\infty}^2)$\\
  \hline

$t_4$ & $\frac{120 (360 e_1+96 e_2+27 e_3+25 e_4+64 e_5+18 e_6)}{-\ell ^4+72 e_1+12 e_2+2 e_4+8 e_5}$ & $-\fft{1260f_{\infty}^2\mu}{1-3f_{\infty}^2 \mu}$\\
  \hline
\end{tabular}
\caption{\small \it  The parameters $\tilde{a}$, $\mathcal{C}_T$ and $t_4$ for the generic massless cubic gravities and Einsteinian cubic gravity in $D=4$. }
\end{center}
\end{table}

Furthermore, we found that the physical requirement $C_T>0$ and $\langle\mathcal{E}\rangle\geq0$ should impose the constraints for coupling constants to take their values within certain appropriate region, both in $D=5$ and $D=4$. Nevertheless, viewing $e_i$s as very small quantities is satisfactory and safe and it would not violate the constraints. Then we also calculated the holographic R\'enyi entropy up to the first order and shear-viscosity-entropy-ratio up to the second order in $D=5$ and $D=4$ respectively. We found the first order effect produced by the higher-order coupling constants $e_i$ on the holographic R\'enyi entropy with taking the limit $q\rightarrow 1$, $q\rightarrow 0$ and $q\rightarrow \infty$ can be indeed expressed by $(a,c,t_4)$ in $D=5$ and $(\tilde{a},\mathcal{C}_T,t_4)$ in $D=4$ in different ways respectively. For shear-viscosity-entropy-ratio, we found in $D=5$, the first order deviation from the Einstein gravity, i.e. KSS bound can be uniquely controlled by $(c-a)/c$ and $t_4$, while up to the second order, the controlling pattern is far from clear; surprisingly, in $D=4$, the deviation up to the second order can even be uniquely controlled by $(\mathcal{C}_T-\tilde{a})/\mathcal{C}_T$ and $t_4$. It should be commented that the discussions of holographic hydrodynamics and R\'enyi entropy should be more involved. In this paper, we aim to shed a light on the controlling pattern of R\'enyi entropy and shear-viscosity-entropy-ratio with respect to universal parameters of unitary CFT, while, e.g. the plasma stability, phase transition which are related to the stability of black holes, and even the superluminal problem are not undertaken. Indeed, in higher order gravities, the black holes are more likely to be unstable in a variety of ways, for example, the perturbation around the black holes would give rise to the Ostrogradsky ghosts \cite{Chen:2013aha,Woodard:2015zca}, the pathological quasi-normal modes \cite{Myers:2010jv} etc (however, these instabilities shall not be mixed with the unitarity of the dual CFT). To fix the instability problem, one has to add more constraints for the coupling constants, e.g. \cite{Myers:2010jv}. Therefore further investigations on the stability of black holes in massless cubic gravities are required in the future such that more rigorous constraints for massless cubic gravities as holographic models can be provided.

\section*{Acknowledgement}
We are grateful to Hong L\"u and Wei-Jian Geng for useful discussions. This work is supported in part by NSFC grants No.~11875200 and No.~11475024.

\appendix

\section{The second order approximate planar black holes}
\label{2order}
In this Appendix, we present the solutions of second order approximate black holes (i.e. $k=0$) in $D=5$ and $D=4$ respectively for the massless cubic gravities.

In $D=5$, $d=4$, the black holes are obtained with $f$ given by
\bea
f(r) &=& \frac{(r^4-r_0^4)}{3 r^{18} \ell _0^{10}} (3 \ell _0^8 r^{16}+2 (e_4 r^8+4 e_5 r^8+252 e_3 r_0^4 r^4+200 e_4 r_0^4 r^4+748 e_5 r_0^4 r^4+172 e_6 r_0^4 r^4
\cr && -108 e_3 r_0^8-97 e_4 r_0^8-320 e_5 r_0^8-68 e_6 r_0^8+60 e_1 (r^8+53 r_0^4 r^4-34 r_0^8)+e_2 (8 r^8
\cr && +844 r_0^4 r^4-452 r_0^8)) \ell _0^4 r^8+4 (e_4^2 r^{16}+16 e_5^2 r^{16}+8 e_4 e_5 r^{16}-151704 e_3^2 r_0^4 r^{12}
\cr && -84523 e_4^2 r_0^4 r^{12}-1472608 e_5^2 r_0^4 r^{12}-69544 e_6^2 r_0^4 r^{12}-227790 e_3 e_4 r_0^4 r^{12}
\cr && -945960 e_3 e_5 r_0^4 r^{12}-713230 e_4 e_5 r_0^4 r^{12}-205440 e_3 e_6 r_0^4 r^{12}-154462 e_4 e_6 r_0^4 r^{12}
\cr && -640120 e_5 e_6 r_0^4 r^{12}-91224 e_3^2 r_0^8 r^8-45517 e_4^2 r_0^8 r^8-948576 e_5^2 r_0^8 r^8-42024 e_6^2 r_0^8 r^8
\cr && -130614 e_3 e_4 r_0^8 r^8-589896 e_3 e_5 r_0^8 r^8-427238 e_4 e_5 r_0^8 r^8-123840 e_3 e_6 r_0^8 r^8
\cr && -88886 e_4 e_6 r_0^8 r^8-399896 e_5 e_6 r_0^8 r^8-648360 e_3^2 r_0^{12} r^4-413113 e_4^2 r_0^{12} r^4
\cr && -6017728 e_5^2 r_0^{12} r^4-286424 e_6^2 r_0^{12} r^4-1039866 e_3 e_4 r_0^{12} r^4-3952248 e_3 e_5 r_0^{12} r^4
\cr && -3181306 e_4 e_5 r_0^{12} r^4-861504 e_3 e_6 r_0^{12} r^4-692106 e_4 e_6 r_0^{12} r^4-2621416 e_5 e_6 r_0^{12} r^4
\cr && +255096 e_3^2 r_0^{16}+181662 e_4^2 r_0^{16}+2266240 e_5^2 r_0^{16}+110632 e_6^2 r_0^{16}+432930 e_3 e_4 r_0^{16}
\cr && +1520952 e_3 e_5 r_0^{16}+1295402 e_4 e_5 r_0^{16}+335904 e_3 e_6 r_0^{16}+285986 e_4 e_6 r_0^{16}
\cr && +999080 e_5 e_6 r_0^{16}+3600 e_1^2 (r^{16}-5571 r_0^4 r^{12}-2671 r_0^8 r^8-34395 r_0^{12} r^4+16883 r_0^{16})
\cr && +8 e_2^2 (8 r^{16}-197273 r_0^4 r^{12}-108713 r_0^8 r^8-1007255 r_0^{12} r^4+448209 r_0^{16})
\cr && +2 e_2 (-8 (6 e_3 (10269 r^{12}+5951 r_0^4 r^8+47861 r_0^8 r^4-20086 r_0^{12})+e_6 (41783 r^{12}
\cr && +24295 r_0^4 r^8+190857 r_0^8 r^4-79531 r_0^{12})) r_0^4+4 e_5 (8 r^{16}-386143 r_0^4 r^{12}-233675 r_0^8 r^8
\cr && -1760461 r_0^{12} r^4+723125 r_0^{16})+e_4 (8 r^{16}-365219 r_0^4 r^{12}-198935 r_0^8 r^8-1822481 r_0^{12} r^4
\cr && +806397 r_0^{16}))+30 e_1 (e_4 (4 r^{16}-87697 r_0^4 r^{12}-45045 r_0^8 r^8-477963 r_0^{12} r^4+222811 r_0^{16})
\cr && +4 e_2 (8 r^{16}-94583 r_0^4 r^{12}-49203 r_0^8 r^8-528645 r_0^{12} r^4+247349 r_0^{16})-4 ((3 e_3 (10003 r^{12}
\cr && +5567 r_0^4 r^8+50697 r_0^8 r^4-22457 r_0^{12})+e_6 (20383 r^{12}+11395 r_0^4 r^8+101117 r_0^8 r^4
\cr && -44541 r_0^{12})) r_0^4+e_5 (-4 r^{16}+94534 r_0^4 r^{12}+55390 r_0^8 r^8+468538 r_0^{12} r^4
\cr && -203130 r_0^{16})))))+\mathcal{O}(e_i^3)\,,\label{f-k=0-D=5-2order}
\eea
and $h$ given by
\bea
h(r) &=& \frac{1}{3 r^{18} \ell _0^{10}}(3 (r^4-r_0^4) \ell _0^8 r^{16}+2 ((60 e_1+8 e_2+e_4+4 e_5) r^{12}+(3120 e_1+836 e_2+252 e_3
\cr && +199 e_4+744 e_5+172 e_6) r_0^4 r^8-3 (2580 e_1+744 e_2+240 e_3+183 e_4+716 e_5+168 e_6) r_0^8 r^4
\cr && +(4560 e_1+1388 e_2+468 e_3+349 e_4+1400 e_5+332 e_6) r_0^{12}) \ell _0^4 r^8
\cr && +4 ((60 e_1+8 e_2+e_4+4 e_5){}^2 r^{20}-2 (10029600 e_1^2+15 (378364 e_2+120036 e_3+87701 e_4
\cr && +378152 e_5+81532 e_6) e_1+789124 e_2^2+75852 e_3^2+42262 e_4^2+736312 e_5^2+34772 e_6^2
\cr && +113895 e_3 e_4+472980 e_3 e_5+356619 e_4 e_5+102720 e_3 e_6+77231 e_4 e_6+320060 e_5 e_6
\cr && +e_2 (492912 e_3+365227 e_4+1544604 e_5+334264 e_6)) r_0^4 r^{16}
\cr && -4 (60 e_1+8 e_2+e_4+4 e_5) (2580 e_1+744 e_2+240 e_3+183 e_4+716 e_5+168 e_6) r_0^8 r^{12}
\cr && +2 (2580 e_1+744 e_2+240 e_3+183 e_4+716 e_5+168 e_6) (3120 e_1+836 e_2+252 e_3
\cr && +199 e_4+744 e_5+172 e_6) r_0^8 r^{12}-(3120 e_1+836 e_2+252 e_3+199 e_4+744 e_5
\cr && +172 e_6) (4560 e_1+1388 e_2+468 e_3+349 e_4+1400 e_5+332 e_6) r_0^{12} r^8-(284781600 e_1^2
\cr && +60 (2529716 e_2+758412 e_3+577435 e_4+2355336 e_5+508132 e_6) e_1+19974944 e_2^2
\cr && +1728000 e_3^2+1046009 e_4^2+16148992 e_5^2+774912 e_6^2+2705292 e_3 e_4+10572336 e_3 e_5
\cr && +8319204 e_4 e_5+4 (578304 e_3+453721 e_4+1765348 e_5) e_6+4 e_2 (2956824 e_3+2282075 e_4
\cr && +9115812 e_5+1980104 e_6)) r_0^{12} r^8+(734799600 e_1^2+480 (824957 e_2+250041 e_3
\cr && +188431 e_4+781001 e_5+167841 e_6) e_1+52658944 e_2^2+4652928 e_3^2+2758609 e_4^2
\cr && +44065328 e_5^2+2093952 e_6^2+7209288 e_3 e_4+28658976 e_3 e_5+22316208 e_4 e_5
\cr && +8 (780024 e_3+605503 e_4+2398004 e_5) e_6
+8 e_2 (3939048 e_3+3009383 e_4
\cr && +12218628 e_5+2642768 e_6)) r_0^{16} r^4-2 (215607600 e_1^2+15 (7779964 e_2+2368164 e_3
\cr && +1776269 e_4+7418312 e_5+1590172 e_6) e_1+15582804 e_2^2+1388124 e_3^2+815364 e_4^2
\cr && +13228040 e_5^2+625092 e_6^2+2140959 e_3 e_4+8576916 e_3 e_5+6648499 e_4 e_5+1862304 e_3 e_6
\cr && +1438863 e_4 e_6+5744156 e_5 e_6+e_2 (9363408 e_3+7120803 e_4+29131484 e_5
\cr && +6284184 e_6)) r_0^{20}))+\mathcal{O}(e_i^3)\,.\label{h-k=0-D=5-2order}
\eea

In $D=4$, $d=3$, we have
\bea
f(r) &=& -\frac{1}{24 r^{13} \ell _0^{10}}(-32 (36 e_1+6 e_2+e_4+4 e_5)^2 r^{15}-24 (r^3-r_0^3) \ell _0^8 r^{12}+(10949904 e_1^2
\cr && +36 (182712 e_2+60993 e_3+46895 e_4+152912 e_5+40662 e_6) e_1+963180 e_2^2+96228 e_3^2
\cr && +63286 e_4^2+614704 e_5^2+42768 e_6^2+159165 e_3 e_4+485028 e_3 e_5+399044 e_4 e_5+128304 e_3 e_6
\cr && +106110 e_4 e_6+323352 e_5 e_6+6 e_2 (103761 e_3+82265 e_4+260696 e_5+69174 e_6)) r_0^3 r^{12}
\cr && +64 (36 e_1+6 e_2+e_4+4 e_5) (360 e_1+96 e_2+27 e_3+25 e_4+64 e_5+18 e_6) r_0^6 r^9
\cr && -24 (360 e_1+96 e_2+27 e_3+25 e_4+64 e_5+18 e_6) (324 e_1+102 e_2+36 e_3+29 e_4+68 e_5
\cr && +24 e_6) r_0^6 r^9+6 (324 e_1+102 e_2+36 e_3+29 e_4+68 e_5+24 e_6) (1044 e_1+246 e_2+54 e_3
\cr && +59 e_4+164 e_5+36 e_6) r_0^9 r^6+8 (5561136 e_1^2+36 (84072 e_2+24435 e_3+21239 e_4+63392 e_5
\cr && +16290 e_6) e_1+406980 e_2^2+32076 e_3^2+26038 e_4^2+220624 e_5^2+14256 e_6^2+58617 e_3 e_4
\cr && +166428 e_3 e_5+150884 e_4 e_5+42768 e_3 e_6+39078 e_4 e_6+110952 e_5 e_6+6 e_2 (38691 e_3
\cr && +34265 e_4+100376 e_5+25794 e_6)) r_0^9 r^6-4 (4 (36 e_1+6 e_2+e_4+4 e_5) r^9+3 (324 e_1
\cr && +102 e_2+36 e_3+29 e_4+68 e_5+24 e_6) r_0^3 r^6-6 (360 e_1+96 e_2+27 e_3+25 e_4+64 e_5
\cr && +18 e_6) r_0^6 r^3+(1044 e_1+246 e_2+54 e_3+59 e_4+164 e_5+36 e_6) r_0^9) \ell _0^4 r^6-2 (42762816 e_1^2
\cr && +36 (642660 e_2+185031 e_3+161767 e_4+484600 e_5+123354 e_6) e_1+3089304 e_2^2
\cr && +236925 e_3^2+195416 e_4^2+1680608 e_5^2+105300 e_6^2+436617 e_3 e_4+1254636 e_3 e_5
\cr && +1143172 e_4 e_5+315900 e_3 e_6+291078 e_4 e_6+836424 e_5 e_6+6 e_2 (290331 e_3+258793 e_4
\cr && +763408 e_5+193554 e_6)) r_0^{12} r^3+(30068496 e_1^2+36 (434616 e_2+117153 e_3+108679 e_4
\cr && +312208 e_5+78102 e_6) e_1+2008332 e_2^2+132678 e_3^2+124808 e_4^2+1003184 e_5^2+58968 e_6^2
\cr && +261657 e_3 e_4+727812 e_3 e_5+702724 e_4 e_5+176904 e_3 e_6+174438 e_4 e_6+485208 e_5 e_6
\cr && +6 e_2 (176121 e_3+166825 e_4+473944 e_5+117414 e_6)) r_0^{15})+\mathcal{O}(e_i^3)\,,\label{f-k=0-D=4-2order}
\eea
and
\bea
h(r) &=& \frac{(r^3-r_0^3)}{24 r^{13} \ell _0^{10}} (24 \ell _0^8 r^{12}+32 e_4^2 r^{12}+512 e_5^2 r^{12}+256 e_4 e_5 r^{12}-96228 e_3^2 r_0^3 r^9-63254 e_4^2 r_0^3 r^9
\cr && -614192 e_5^2 r_0^3 r^9-42768 e_6^2 r_0^3 r^9-159165 e_3 e_4 r_0^3 r^9-485028 e_3 e_5 r_0^3 r^9-398788 e_4 e_5 r_0^3 r^9
\cr && -128304 e_3 e_6 r_0^3 r^9-106110 e_4 e_6 r_0^3 r^9-323352 e_5 e_6 r_0^3 r^9-49572 e_3^2 r_0^6 r^6-36078 e_4^2 r_0^6 r^6
\cr && -476592 e_5^2 r_0^6 r^6-22032 e_6^2 r_0^6 r^6-87885 e_3 e_4 r_0^6 r^6-324324 e_3 e_5 r_0^6 r^6-276420 e_4 e_5 r_0^6 r^6
\cr && -66096 e_3 e_6 r_0^6 r^6-58590 e_4 e_6 r_0^6 r^6-216216 e_5 e_6 r_0^6 r^6+4 (4 e_4 r^6+16 e_5 r^6+108 e_3 r_0^3 r^3
\cr && +91 e_4 r_0^3 r^3+220 e_5 r_0^3 r^3+72 e_6 r_0^3 r^3-216 e_3 r_0^6-167 e_4 r_0^6-380 e_5 r_0^6-144 e_6 r_0^6
\cr && +36 e_1 (4 r^6+31 r_0^3 r^3-47 r_0^6)+6 e_2 (4 r^6+55 r_0^3 r^3-95 r_0^6)) \ell _0^4 r^6-971028 e_3^2 r_0^9 r^3
\cr && -684128 e_4^2 r_0^9 r^3-6200528 e_5^2 r_0^9 r^3-431568 e_6^2 r_0^9 r^3-1658637 e_3 e_4 r_0^9 r^3-4849956 e_3 e_5 r_0^9 r^3
\cr && -4123348 e_4 e_5 r_0^9 r^3-1294704 e_3 e_6 r_0^9 r^3-1105758 e_4 e_6 r_0^9 r^3-3233304 e_5 e_6 r_0^9 r^3
\cr && +1560060 e_3^2 r_0^{12}+1077620 e_4^2 r_0^{12}+10193840 e_5^2 r_0^{12}+693360 e_6^2 r_0^{12}+2642463 e_3 e_4 r_0^{12}
\cr && +7914348 e_3 e_5 r_0^{12}+6667132 e_4 e_5 r_0^{12}+2080080 e_3 e_6 r_0^{12}+1761642 e_4 e_6 r_0^{12}
\cr && +5276232 e_5 e_6 r_0^{12}+1296 e_1^2 (32 r^{12}-8417 r_0^3 r^9-6441 r_0^6 r^6-96923 r_0^9 r^3+152213 r_0^{12})
\cr && +36 e_2^2 (32 r^{12}-26723 r_0^3 r^9-18123 r_0^6 r^6-288605 r_0^9 r^3+459131 r_0^{12})+36 e_1 (256 e_5 r^{12}
\cr && -60993 e_3 r_0^3 r^9-152656 e_5 r_0^3 r^9-40662 e_6 r_0^3 r^9-41553 e_3 r_0^6 r^6-119568 e_5 r_0^6 r^6
\cr && -27702 e_6 r_0^6 r^6-646785 e_3 r_0^9 r^3-1626784 e_5 r_0^9 r^3-431190 e_6 r_0^9 r^3+1034451 e_3 r_0^{12}
\cr && +2626240 e_5 r_0^{12}+689634 e_6 r_0^{12}+24 e_2 (16 r^{12}-7597 r_0^3 r^9-5529 r_0^6 r^6-84394 r_0^9 r^3
\cr && +133576 r_0^{12})+e_4 (64 r^{12}-46831 r_0^3 r^9-32079 r_0^6 r^6-518611 r_0^9 r^3+817993 r_0^{12}))
\cr && +6 e_2 (-27 (3 e_3+2 e_6) (1281 r^9+785 r_0^3 r^6+13313 r_0^6 r^3-21331 r_0^9) r_0^3+8 e_5 (32 r^{12}
\cr && -32555 r_0^3 r^9-23955 r_0^6 r^6-338069 r_0^9 r^3+548123 r_0^{12})+e_4 (64 r^{12}-82201 r_0^3 r^9
\cr && -51609 r_0^6 r^6-887197 r_0^9 r^3+1405207 r_0^{12})))+\mathcal{O}(e_i^3)\,.\label{h-k=0-D=4-2order}
\eea
\section{The locality of the excitation operator}
\label{local}
In this Appendix, we briefly review why the excitation operator $\mathcal{O}$ like $T_{ij}\varepsilon^{ij}$ discussed in section \ref{flux} is localized at $\rho=\ell$ and $y^1=y^2=0$. We start with (\ref{coordinate-transf}) and (\ref{AdS-coor}), for clear, we may present them here as well
\bea
ds^2 &=& \fft{\ell^2}{r^2}dr^2+r^2 \eta_{ij}dx^i dx^j
\cr &&
\cr &=& \fft{\ell^2}{r^2}dr^2+r^2 (-dx^{+}dx^{-}+dx^{\tilde{i}}dx_{\tilde{i}})
\cr &&
\cr &=& \fft{\ell^2}{\rho^2}d\rho^2+\rho^2 (-dy^{+}dy^{-}+dy^{\tilde{i}}dy_{\tilde{i}})\,,\label{AdS-coor-appen}
\eea
in which we have
\bea
y^{+}=-\fft{1}{x^{+}}\,,\qquad y^{-}=x^{-}-\fft{x_{\tilde{i}} x^{\tilde{i}}}{x^{+}}-\fft{\ell^2}{r^2 x^{+}}\,,\qquad y^{\tilde{i}}=\fft{x^{\tilde{i}}}{x^{+}}\,,\qquad \rho=r x^{+}\,.\label{coordinate-transf-appen}
\eea
For our purpose, it is essential to recall the embedding picture of AdS, i.e. AdS can be defined by embedding itself in a higher dimensional space with the signature as $(-1,-1,1,\cdots)$, explicitly
\be
ds_{d+2}^2=-(dX^{-1})^2-(dX^0)^2+\sum_{a=1}^{d}(dX^a)^2\,,\qquad -(X^{-1})^2-(X^0)^2+
\sum_{a=1}^{d}(X^a)^2=-\ell^2\,.\label{emb}
\ee
We now have
\bea
&& r=X^{-1}+X^d\,,\qquad x^i=\fft{\ell X^i}{r}\,,
\cr && \rho=X^0+X^{d-1}\,,\qquad y^{+}=-\fft{\ell}{\rho}(X^{-1}+X^d)\,,\qquad
y^{-}=-\fft{\ell}{\rho}(X^{-1}-X^d)\,,
\cr && y^{\tilde{i}}=\fft{\ell X^{\tilde{i}}}{\rho}\,,\qquad \tilde{i}=1,\cdots,d-2\,.\label{emb-pic}
\eea
Keep (\ref{emb-pic}) in mind, we are in the right position to consider the bulk field corresponding to the excitation operator $\mathcal{O}$ with the scaling dimension $\Delta$.
\be
\phi(x,r)=\int d^4 x'\fft{r^{-\Delta}\ell^{\Delta}}{\Big((x-x')^2+r^{-2}\ell^2\Big)^{\Delta}}\phi_0(x)\,,
\qquad \phi_0(x)\sim e^{-i p\cdot x}\,.\label{prop}
\ee
Note primarily we should take $r\rightarrow\infty$ in (\ref{prop}), then by using (\ref{emb}) we come to
\be
\lim_{r\rightarrow\infty}\fft{(x-x')^2 r}{\ell}=-\fft{2\ell}{r}\eta_{ij}X^i X'^j+\fft{2\ell}{r}(X^{-1}X'^{-1}-X^d X'^d)=-\fft{2\ell}{r}X\cdot X'\,,
\ee
where we make use of the fact that $X^{-1}$ and $X^d$ is not dependent on $x$, consequently they shall make no difference with $X'^{-1}$ and $X'^d$. We then have
\be
\phi\sim\int d^4 x'\fft{r^{\Delta}\ell^{-\Delta}}{(X\cdot X')^{\Delta}}\phi_0(x)\,,
\qquad \phi_0(x)\sim e^{-i p\cdot x}\,.\label{prop2}
\ee
Note the energy flux is measured in the surface of $y^{+}=0$ which implies $X^{+}=X^{-1}+X^d=0$, hence we have
\be
\fft{\ell}{r}X\cdot X'=\eta_{ij}X^i x^j-\fft{\ell}{2} X^{-}\,,\qquad X^{-}=X^{-1}-X^d\simeq2X^{-1}\,.
\ee
The resulting propagator is thus given by
\be
\phi\sim\int d^4 x'\fft{e^{-i p\cdot x}}{(-X^0t+X^{\tilde{i}}x_{\tilde{i}}-\fft{\ell}{2}X^{-})^{\Delta}}\,.\label{prop3}
\ee
For simplicity, we focus on the transverse mode, i.e. $p=(E,0,\cdots,0)$. Then, we integrate (\ref{prop3}) over $t$, slipping off inessential numerical factors, we obtain
\be
\phi\sim\int d^3 x' \fft{(E)^{\Delta-1}}{(X^0)^\Delta}e^{i(X^{\tilde{i}}x_{\tilde{i}}-
\fft{\ell}{2}X^{-})\fft{E}{X^0}}\,.
\ee
Slipping off all factors that are irrelevant to the localized property of the excitation operator, we end up with
\be
\phi(X^{+}=0,X^{-},X^i)\sim e^{-i\fft{\ell}{2}EX^{-}/X^0}\delta^3(X^{\tilde{i}})\,.\label{local-1}
\ee
It is now evidently to see from (\ref{local-1}) that the perturbation is localized at $X^{\tilde{i}}=0$. Transforming to $y$-coordinates, $X^{\tilde{i}}=0$ implies $y^1, y^2,\cdots ,y^{d-2}=0$ and $\rho=X^0$. From the embedding picture (\ref{emb}), we now should have $X^0=\ell$, which immediately suggests that $\rho$ is localized at $\rho=\ell$. To be precise, we have
\be
\phi(y^{+}=0,y^{-},y^{1},\cdots,y^{d-2},\rho)\sim e^{i \fft{E}{2\ell}y^{-}} \delta(y^1)\cdots \delta(y^{d-2})\delta(\rho-\ell)\,.
\ee
Therefore, holographically, the operator is localized at $\rho=\ell$, $y^1=y^2=0$.

\end{document}